\documentclass[amssymb, nobibnotes, aps, pra,twocolumn]{revtex4-2}
\usepackage{graphicx}
\usepackage{amsmath,mathrsfs,dcolumn,dsfont}
\usepackage{amssymb}
\usepackage{epsfig,epstopdf}
\usepackage{bm,bbm,mathrsfs}
\usepackage{threeparttable}
\usepackage{comment}
\usepackage{ulem}
\usepackage{tikz}
\usepackage[percent]{overpic}
\usepackage[large]{subfigure}
\usepackage{color}
\usepackage{xr}
\usepackage{hyperref}
\usepackage{euscript}
\usepackage[top=30truemm,bottom=30truemm,left=26truemm,right=26truemm]{geometry}
\hypersetup{
	colorlinks   = true, 
	urlcolor     = blue, 
	linkcolor    = blue, 
	citecolor   = blue 
}
\begin{document}
	
\title{Incoherent control of two-photon induced optical measurements in open quantum systems: quantum heat engine perspective}
\author{Md Qutubuddin and Konstantin E. Dorfman}
\email{dorfmank@lps.ecnu.edu.cn}
\affiliation{State Key Laboratory of Precision Spectroscopy, East China Normal University, Shanghai 200062, China}%

\date{\today}
\begin{abstract}
We present a consistent optimization procedure for the optical measurements in open quantum systems using recently developed incoherent control protocol. Assigning an effective hot bath for the two-entangled-photon pump we recast the transmission of classical probe as a work in a quantum heat engine framework. We demonstrate that maximum work in such a heat engine can exceed that for the classical two-photon and one-photon pumps, while efficiency at maximum power can be attributed to conventional boundaries obtained for three-level maser heat engine. Our results pave the way for incoherent control and optimization of optical measurements in open quantum systems that involve two-photon processes with quantum light.
\end{abstract}
\pacs{Valid PACS appear here}
\maketitle
\section{Introduction}
Thermal engine \cite{c24} plays a pivotal role in many thermodynamical processes. In the seminal work by Scovil and Schulz-DuBois \cite{sb59}, a maser heat engine was formulated in the context of the detailed balance which results in the maximum efficiency \cite{sbj59, sb67} for a three level system operating between hot and cold baths. The ascent of quantum heat engines (QHEs) attracted a significant amount of attention in the last few decades and constitutes an important research direction within the quantum thermodynamics, both theoretically \cite{bh13,bm17,bc19, rq84, jmc16,kbn21,sga21,bg17,ud21} and experimentally \cite{kmp14,jdt16,kf17,pb19,lgs19, kbl19}. In addition to the diverse development of QHEs \cite{nba21, tas18, msm21}, it has intrinsic relationship with the real physical systems such as laser, solar cell\cite{sd11, gg18}, battery \cite{cyc21,mbkb}, light harvesting \cite{dv13}, etc. While some of the promising features  of QHEs such as quantum coherence and entanglement \cite{sprl, sz03, rm12, lk15} show a possibility for enhancing the maximum output power for resonantly driven systems \cite{osm20}, the significance of entanglement in optical measurements in open quantum systems from the QHE perspective has not been investigated so far. 

Recently, authors developed an incoherent control method of optical signals \cite{qd21} that views the pump-probe measurements as a QHE, which transfers energy from the pump pulse to the probe pulse treating the dissipation to the environment explicitly, while computing the work performed by the system via the detected probe photons. In this method we have introduced an effective thermal bath by combining a coherent pump pulse excitation of electronic excited state of molecule with the thermal relaxation. The “incoherent” control algorithm for the optical signals in open quantum systems is then introduced based on the analogy with the QHE. It has been further shown that the spectroscopic measurement for the probe pulse transmission can be improved when the corresponding parameter regime is close to the limiting operation such as the Curzon-Ahlborn limit etc.

In the course of the above developments we realized that the efficient operation of QHE in the open quantum systems such as molecules is strongly correlated with the ability to efficiently excite a particular electronic state. In the same time the various molecular degrees of freedom such as nuclear motion, the complex selection rules, and the associated dissipation processes restrict the ability to control the excited states probabilities. Moreover the latter are governed by the uncertainty relation between the spectral bandwidth of the molecules and the temporal profile of the excitation pulse. It has been further shown that the two photon excitation with the entangled photons can efficiently control the multi exciton population distribution in the complex molecules since these can violate the uncertainty principle in the two-photon absorption measurements\cite{sdf13}. In addition to the spectral selectivity, the entangled two-photon absorption probability scales linearly with the pump intensity in contrast with the classical two-photon absorption which is quadratic in the pump intensity \cite{ryangood2020}. This feature makes it attractive for the low intensity applications in photo sensitive materials such as biological molecules etc \cite{smuk20road}. While a significant amount of experimental \cite{zhpag2021} effort has been dedicated for the optimization of two-photon absorption measurement using quantum light, it lacks a formalized theoretical foundation due to the dissipative nature of the open quantum system. It is therefore imperative to develop a consistent optimization procedure for such optical measurements in connection with the fundamental frameworks such as QHEs. 

In this article we explore the QHEs analogy with the setup that involves two-photon pump excitation with both classical and quantum states of light by utilizing the  effective thermal bath \cite{qd21}. The two photons can be initially in e.g., entangled twin photon state, which allows us to explore the effects of entanglement in QHE operation. By presenting a consistent technique of maximization of power and efficiency at maximum power for the two-photon pumped QHE using incoherent control, one can manipulate the two-photon induced fluorescence (TPIF) and pump-probe signals due to additional control parameter (entanglement time) \cite{dsm16, sdf13} which does not exist classically.

\section{Effective Heat Bath}
\label{effective heat bath}
We consider a three-level molecular system with ground state $g$, single excited electronic state $e$ and double excited electronic state $f$ (see in Fig. 1). To keep the notations consistent with our previous work \cite{qd21} we denote vibrational states of electronic ground state as $0$ and $g$, while vibrational states of double excited electronic state are $2$ and $1$, and $e$ and $e'$ denote vibrational states of single excited electronic state.
\begin{figure*}
	\includegraphics[width=0.9\textwidth]{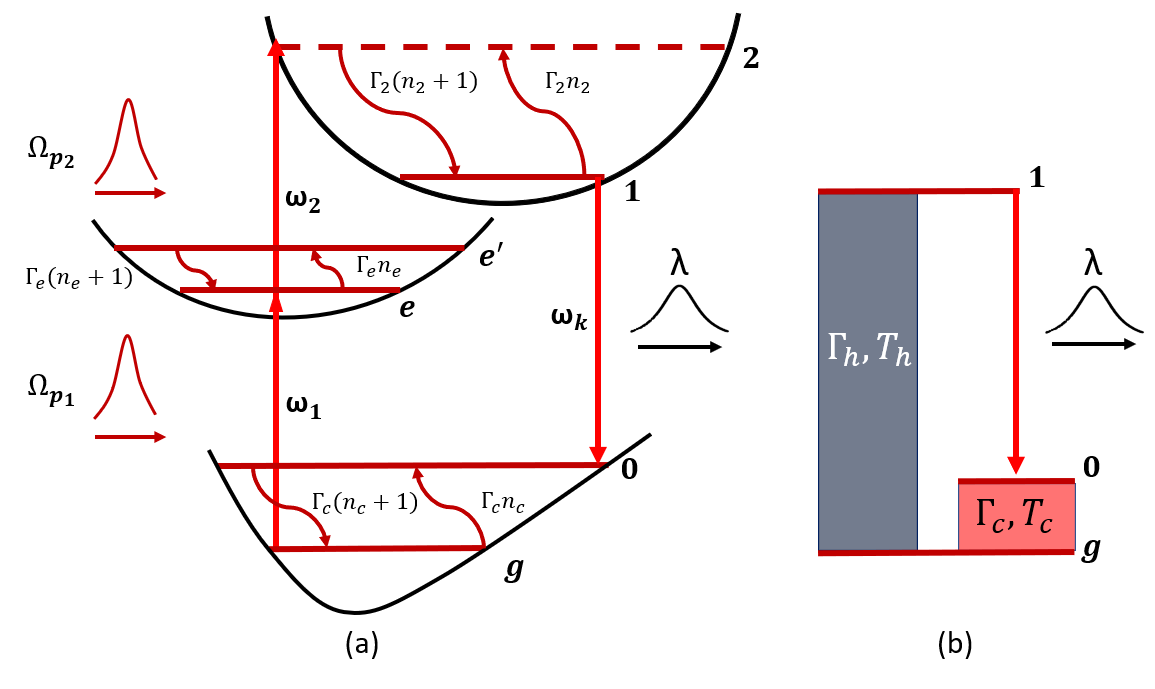}
	\caption{(a) Schematic for the three-level molecule undergoing two-photon pump - probe measurements.  The pump field resonant with transition $g-2$ excites a vibrational wave packet in the higher vibrational state $2$ via the intermediate levels $e$ and $e'$, which relaxes to the lower energy vibrational state $1$. The probe field then stimulates the emission from the state $1$ to the excited vibrational level $0$ in the ground electronic state. Finally, vibrational relaxation brings the system back to its ground state $g$. (b) Equivalent three-level QHE with transitions between energy levels $g - 1$ and $g - 0$ driven by hot (at temperature $T_h$) and cold (at $T_c$) heat baths. The single-mode stimulated emission representing the work done by the QHE occurs at $1 - 0$ transition with the Rabi frequency $\lambda$.} \label{qsys}
\end{figure*}
A two-photon pump field excites the molecule from $g$ to $2$ via virtual states $e-e'$ with the pump Rabi frequencies $ \Omega_{i}$ and central frequencies $\omega_i$, $i=1,2$. The vibrational state $2$ relaxes to $1$ by emission of phonon. The stimulated emission $1-0$ due to interaction with probe field of Rabi frequency $\lambda$ followed by the thermal relaxation via interaction of $0-g$ transition with the cold bath which then brings the system to its initial ground states.
 The total Hamiltonian of the system is given by
 \begin{align}
 \hat{H}_{tot.} = \hat{H}_{0} + \hat{H}_{I} + \hat{H}_{I.V.}, \label{h1}
 \end{align}
where subscript $I$ indicates the light-matter interaction of pump and probe fields and $I.V.$ indicates the interaction with the vibrational modes. $\hat{H}_{0} = \sum_{i} \omega_{i}|i \rangle\langle i| $ where $i= g,0, e, e', 1,2$. The pump-molecule interaction Hamiltonian in the rotating wave approximation reads
\begin{align}
\hat{H}_{I}(t) = -\hat{V}^{\dagger}(t) \sum_{i=1,2} (\hat{E}_{j}(t) + \hat{E}^{\dagger}_{j}(t)),
\end{align}
where $\hat{V}(t) = \mu_{ge} \ e^{-i\omega_{e}t} \ |g\rangle\langle e| + \mu_{e2} \  e^{-i \omega_2 t} \ |e \rangle \langle 2 |$, and  $\hat{E}_{j}(t) = i \int_{0} ^{\infty} d\omega_{j} \sqrt{\frac{\hbar {\omega}_{j} } {2\mathbf{V}\epsilon_{0} }} \hat{a}_{j}(\omega_{j}) e^{-i\omega_j t}$, where $j = 1, 2,$ indices denote the first and second photons, where $\hat{a}_j$ and $\hat{a}^{\dagger}_j$ are the annihilation and creation operators for the $j$-th photon that satisfy the commutation relation $\left[ \hat{a}_j(\omega), \hat{a}_{j'}^{\dagger}(\omega') \right] = \delta (\omega - \omega')\delta_{jj'} $, $\mathbf{V}$ is the quantization volume and $H_{I.V} = \sum_{m, i<j}^{} b^{\dagger}_{m} |i \rangle \langle j| e^{-i\omega_{i}t} $. The two-photon pump utilized in our work originates from the single photon sources so the coupling to the system is naturally weak. In the case of multi photon sources such as squeezed light the coupling can be moderate and yet, it is far from being strong \cite{oleg2020}. Hence, we consider that the couplings to pump fields are weak so we can use the perturbation theory. 

 
Assuming that all molecules are initially in the ground state, the density matrix of interacting matter-field system at time $t$ is given in the interaction picture by the time-ordered exponential superoperator
\begin{align}
{\hat{\rho}}(t) = \hat{\mathcal{T}} \text{exp}\left[- \frac{i}{\hbar} \int^{t}dt' \hat{H} _{int,-} (t') \right] \hat{\rho}_{mat} \otimes \hat{\rho}_{\text{field}},
\end{align} 
where $\mathcal{T}$ is a time ordering superoperator, the interaction Hamiltonian superoperator $H_{int,\pm}$ is defined by its action on the ordinary operator $X$ as: $H_{int, -} X = H_{int} X - X H_{int} ~~ \text{and} ~~H_{int, +} X = H_{int} X + X H_{int}$ \cite{rahmu2010},
\begin{figure*}
\includegraphics[width=0.98\textwidth]{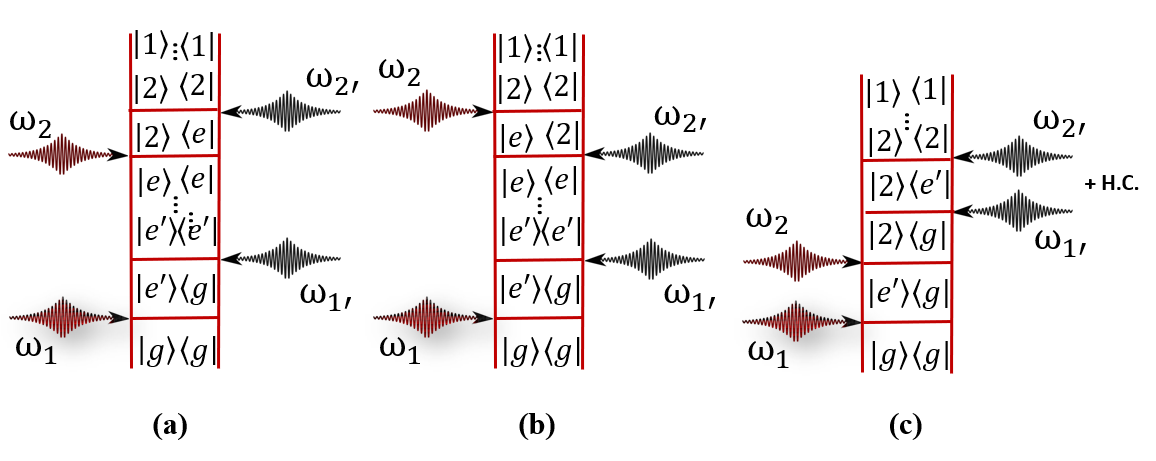}
\caption{(a) The set of double sided Feynman diagrams representing the leading order contribution to the population of excited state due to relaxation of $2 - 1$. There are a total of six pathways that contribute to the matter field system: diagrams (a)–(c), and their complex conjugate.}\label{feyd}
\end{figure*}
$\hat{\rho}_{mat}$ denotes the ground state of the matter system, and $\hat{\rho}_{\text{mat}}$ denotes the initial state of the light field. The leading-order contribution to the population of level $1$ is given by the convolution of four-point field and matter correlation functions using Feynman diagram in Fig. \ref{feyd} is obtained in Eqs. (\ref{eqf1}), (\ref{eqf2}) and (\ref{eqf3}) in Appendix A. The total population of state $1$ due to pumping transition $g - 2$ followed by the relaxation $2 \rightarrow 1$ is given by $\rho_{11}(t) = \rho^{a}_{11}(t) + \rho^{b}_{11}(t) + \rho^{c}_{11} (t)$ obtained in Eq.(\ref{A22}), and it recast as by introducing the detuning $\delta = \omega_{2e} - \omega_{2e'}$ and assuming the pump is tuned midway between $e$ and $e'$ states $\omega_{0} = \frac{1}{2} (\omega_{2e'} + \omega_{2e}) $. The solution we obtain the population of level $1$ from Eq. (\ref{A22}) reads
	
\begin{align}
\rho_{11}(t) &= \frac{16 \, \delta^{2} \, \tilde{\delta}^{2} \, \Omega^{4}_{p} \ (1 - e^{- \Gamma_{2}(2 n_2 + 1)}) }{(2 n_2 + 1) (\delta^{2} + 4 \sigma^{2}_{p})^{2}(\tilde{\delta}^{2} + 4\sigma^{2}_{p} +)^{2}},\nonumber\\ \rho_{gg}(t) &= 1 - \rho_{11}(t), \ \ \label{eq14}
\end{align}
where $\tilde{\delta} = \delta + 2 \omega_{2e'} - 2\omega_{e'g}$. Before proceeding to the QHE model, which is based on the perturbative solution of the complete set of equations given by Eq. (\ref{A1}) (Appendix A), we first introduce an effective heat bath. To that end, we assume that the probe field is much stronger than the coupling to the phonon bath that governs the $2-1$ transition, which itself is stronger than that of the bath driving $0-g$ transition: $\lambda >> \Gamma_{2}n_2 >> \Gamma_c n_c$. The latter condition can be obtained in a variety of molecular systems \cite{harris1989}. Under these conditions one can eliminate the state $0$ from the total system of Eq. (\ref{A1}) and consider only three states such that the coherent excitation $g-2$ is followed by a relaxation $2-1$ in Fig 1b. The combined effect of the coherent excitation $g\rightarrow 2$ followed by the phonon relaxation $2 \rightarrow 1$ can be replaced by an effective thermal bath at temperature $T_h$ with the average photon number $n_h = \left[ \text{exp}(\hbar \omega_{1g}/k_{B}T_{h}) -1 \right] ^{-1}$ and dephasing $\Gamma_h$. In this case state $2$ can be eliminated and the corresponding equation of motion for the populations of $g$ and $1$ read
\begin{align}
\dot{\rho}_{11} = -\Gamma_{h} \left[(n_h + 1)\rho_{11} - n_{h}\rho_{gg} \right], \ \dot{\rho}_{gg} + \dot{\rho}_{11} = 0, \ \ \label{teq1}
\end{align}
which yields the time-dependent solution
\begin{align}
\rho^{th}_{11}(t) = \mathcal{N}_{th} n_{h}(1 - e^{- \Gamma_{h}(2 n_{h} + 1)t}), \ \rho^{th}_{gg} + \rho^{th}_{11} = 1, \ \ \label{teq2}
\end{align}
where superscript $"th"$ indicates the thermal bath and the normalization
$\mathcal{N}_{th} = \left[1 + 2 n_h \right]^{-1}$. Following the approach outlined in Ref. \cite{qd21}, the population of level $1$ excited by a coherent drive is given in Eq. (\ref{eq14}) and the $\rho_{gg}(t)$ is obtained using the population conservation
$\rho_{11}(t) + \rho_{gg}(t) = 1$. This coherently excited populations $1$ and $g$  match with the thermal bath driven populations $1$ and $g$ given in Eq. (\ref{teq2}). By matching the two we obtain $n_h$ and $\Gamma_h$, that must satisfy
\begin{align}
n_{h} &= \frac{16 \ \Omega^{4}_{p} \ (n_2 + 1) \delta^{4}}{(2 n_2 + 1)(\delta^{2} + 4 \sigma^{2}_{p})^{4} - 32 \ \Omega^{4}_{p} \ (n_2 + 1) \ \delta^{4} },\nonumber\\
\Gamma_{h} &= \Gamma_{2} \frac{(2 n_2 + 1)(\delta^{2} + 4 \sigma^{2}_{p}) - 32 \ \sigma^{4}_{p} (n_2 + 1)  \delta^{4} }{(\delta^{2} + 4 \ \sigma^{2}_{p})^{4} },  \label{ppm}
\end{align}
where we set $\omega_{2e'} \simeq \omega_{e'g}$.
The effective thermal bath parameters defined in Eq. (\ref{ppm}) yield the quantitative population match between coherent and thermal baths shown in Fig. \ref{fppm}a. The agreement between the populations shown in Fig. \ref{fppm}b ensures the qualitative formation of an effective bath.
\begin{figure*}
	\includegraphics[width=0.99\textwidth]{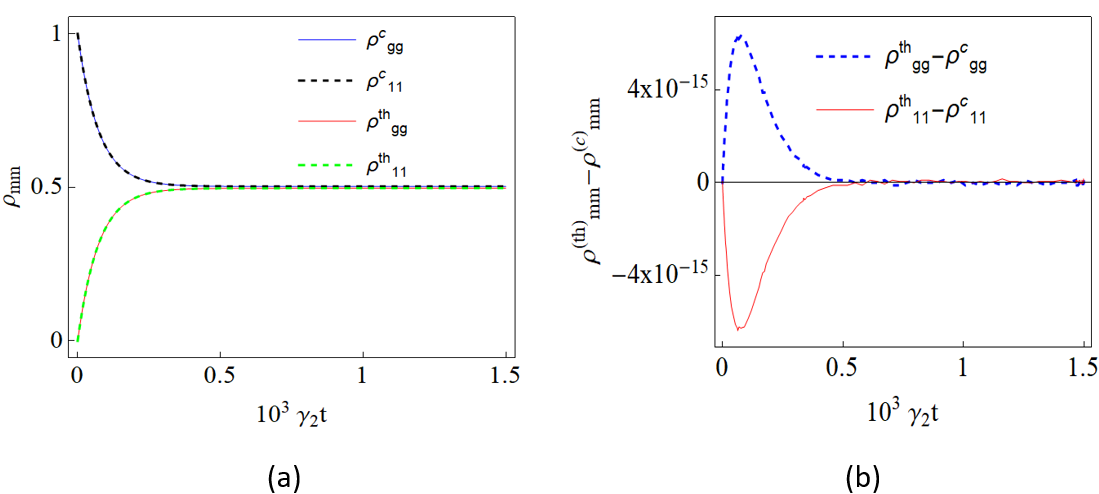}
	\caption{(a)The population of ground $g$ and lowest excited state 1 obtained using the coherent bath in Eq. (\ref{eq14}) (solid lines) and the thermal bath in Eq. (\ref{teq2}) ( dot-dashed) using parameters in Eq. (\ref{ppm}).  (b) The difference between populations of coherent and thermal baths $\rho_c - \rho_{th}$ for parameters in Eq. (\ref{ppm}). The parameters read $n_2 = n_c = 100, \Gamma_{2} = \Gamma_c = 0.002 \text{ps}^{-1}, \Omega_{p} = 0.0078 \text{ev} \  \text{and} \ \sigma_{p} = 30.34 \text{cm}^{-1}.$} \label{fppm}
\end{figure*}

\section{Quantum Heat Engine}
We next obtain for the power  and efficiency given by \cite{kpra18}
\begin{align}\label{eq:Pdef}
	&P= i\hbar\lambda(\omega_{c}-\omega_{h})(\rho_{01}-\rho_{10}),\notag\\
	&\dot{Q}_{h}= i\hbar\omega_{h}\lambda(\rho_{01}-\rho_{10}), \notag\\
	&\eta = 1 - \frac{\omega_{c}}{\omega_{h}}.
\end{align}
where $\omega_{h} = \omega_{1} - \omega_{g}$ and $\omega_{c} = \omega_{0} - \omega_{g}$, $ n_{h} $ and $n_{c}$, $\Gamma_{h} $ and $\Gamma_{c}$ are the average occupation numbers and dephasing rates for the hot and cold baths, respectively. To find the steady state solution for $\varrho_{01}$ and $\varrho_{10}$ in Eq. (\ref{eq:Pdef}), we follow the standard procedure for the three level molecule given in the Supplementary Material of Refs. \cite{sd11} and \cite{qd21}. 
The output power and efficiency for the three level QHE then read
\begin{align}
P &=\frac{2}{3} \frac{\lambda^{2}\Gamma_{h}\Gamma_{c}(n_{c} - n_{h}) (\omega_{c} - \omega_{h}) }{(\Gamma_{h}n_{h} + \Gamma_{c}n_{c}) (\lambda^{2} + \Gamma_{h}\Gamma_{c}n_{c}n_{h} ) }, \nonumber\\ \eta &= 1 - \frac{1}{c_p - c_{21}},\label{genp}
\end{align}
where $c_{p} = \omega_p/\omega_c$ and $c_{21} = \omega_{21}/\omega_c$.
\subsection{Classical-two-photon pump}
We now recast the output power in Eq. (\ref{genp})  using Eq. (\ref{ppm}) for in the high temperature limit where $n_{c} = n_{1} \simeq T_{c}/ \omega_{c}$, $n_{2} =n_{h} \simeq T_{2}/ \omega_{21}$, where $\omega_c = \omega_{0g}$. We then introduce an effective temperature of the hot bath $T_{h} = (\Omega_{p} \ \Gamma^{2}_{2}\ / 2 \ \delta )^{1/2}$ and the dimensionless temperature scale: $\tau = T_c/T_h$. The pump energy scale $c_p = \omega_{p}/ \omega_{c}$, the coupling scale: $\lambda' = \lambda (\Gamma_{2} T_{c})^{-1/2}$ and the pump pulse width scale: $\sigma'_{p} = \sigma^{e}_{p} \ \Gamma_{2} / \delta T_{c}$, where $\sigma^{e}_{p} = (\sigma^{2}_{p} - \delta^{2}/4)^{1/2}$. Eq. (\ref{genp}) for dimensionless parameters given in Eq. (\ref{eq20}) can be finally maximized with respect to dimensionless variable $c_{21}$ which yields
\begin{align}
P^{max}_C = \frac{4 u v \lambda ' \tilde{\tau} \big( 2 \, \EuScript{A} + 2 \,\alpha  u \, v + \tau ^8 \sigma'^{8}_p c'_p (m u + v)  \big)} {3 \tau^8 \sigma'^8_p \left(v - u \lambda '\right)^2}, \ \ \label{eq21}
\end{align}
where $ \EuScript{A} = \sqrt{u \, v \, \left(\tau^8 \, c'_p \,\sigma'^8_p \, + \alpha \, u\right) \left(\tau ^8 \, c'_p \,\lambda ' \,\sigma'^8_p + \alpha  v\right)}$, $\tilde{\tau} = 1 - \tau ^8  \sigma'^{8}_p $ and $ c'_{p} = c_{p} - 1$.
The efficiency corresponding to the maximum output power defined in Eq. (\ref{eq21}) is given by
\begin{align}
\eta^{*}_{C} = 1 - \frac{1}{c_p + \frac{\tilde{\tau} \sqrt{u v \left(c'_p \tau ^8 \sigma'^8_p + \alpha  u \right) (\tau ^8 \lambda' c'_p \sigma'^8_p + \alpha  v )} + u v \, \tilde{\tau}\alpha ^2 } {\tilde{\tau} \, \tau^8 \sigma'^8_p \left( \alpha  v + \lambda ' \left(c'_p \tau ^8 \sigma'^8_p + \alpha  u\right) \right)} },\nonumber \\ \label{eff1}
\end{align}
where subscript $C$ specifies the efficiency of two-photons pump.  We next assume the weak dissipation regime i.e., $\omega_{c}>> \Gamma_{c}$ which yields
\begin{align}
\eta^{*}_{CW} = 1 - \frac{1}{c_p + \frac{\alpha ^2 u v}{\tau ^8 \sigma'^8_p \left(\tau ^8 \left(c_p - 1\right) \lambda' \sigma'^8_p + \alpha  u \lambda'\right)}},\label{eff2}
\end{align}
where subscript $CW$ indicates the classical efficiency in the weak coupling regime.
\begin{table}[h]
	\begin{tabular}{ c c c c }
		\hline\hline
		Bound  & $\eta^*_{CW}$ & $c_p$ & $\sigma'^{C}_{p}$\\[0.2cm] \hline
		I & $0$ & $1$ & $\sqrt[8]{\xi  \left(u-\sqrt{u^{2} - \frac{4 u v}{\lambda '}}\right)} $ \\[0.2cm] 
		I/II & $\frac{\eta_C}{2}$ & $\frac{2}{2-\eta_C}$ & $\sqrt[8]{\xi  \left(u - \sqrt{ u^{2} -  \frac{2 u v \left(2-\eta _C\right) }{ \lambda '}} \right)} $  \\[0.2cm] 
		II/III& $\eta_{CA}$ &  $\frac{1}{\sqrt{1-\eta_C}}$ & $ \sqrt[8]{ \xi  \left(u-\sqrt{ u^{2} - \frac{4 u v \sqrt{1-\eta _C} }{\lambda '}}\right)}$ \\[0.2cm] 
		III/IV & $\frac{\eta_C}{2-\eta_C}$ & $\frac{2-\eta_C}{2(1-\eta_C)}$ & $\sqrt[8]{\xi  \left(u - \sqrt{u^{2} - \frac{8 u v \left(1-\eta _C\right)}{\left(2-\eta _C\right) \lambda '}}\right)} $ \\[0.2cm] 
		IV&$\eta_C$ & $\frac{2}{1-\eta_C}$ & $ \sqrt[8]{\xi  \left(u-\sqrt{ u^{2} - \frac{4 u v \left(1-\eta _C\right)}{\lambda '}}\right)} $ \\[0.2cm] \hline
		\hline \label{tab1}
	\end{tabular}
	\caption{Parameters of the coherent bath corresponding to the QHE efficiency bounds shown in Fig. \ref{et1}, where $\xi = \frac{\alpha}{2 \left(1-\eta _C\right){}^8}$. }
\end{table}
\begin{figure*}
	\includegraphics[width=0.97\textwidth]{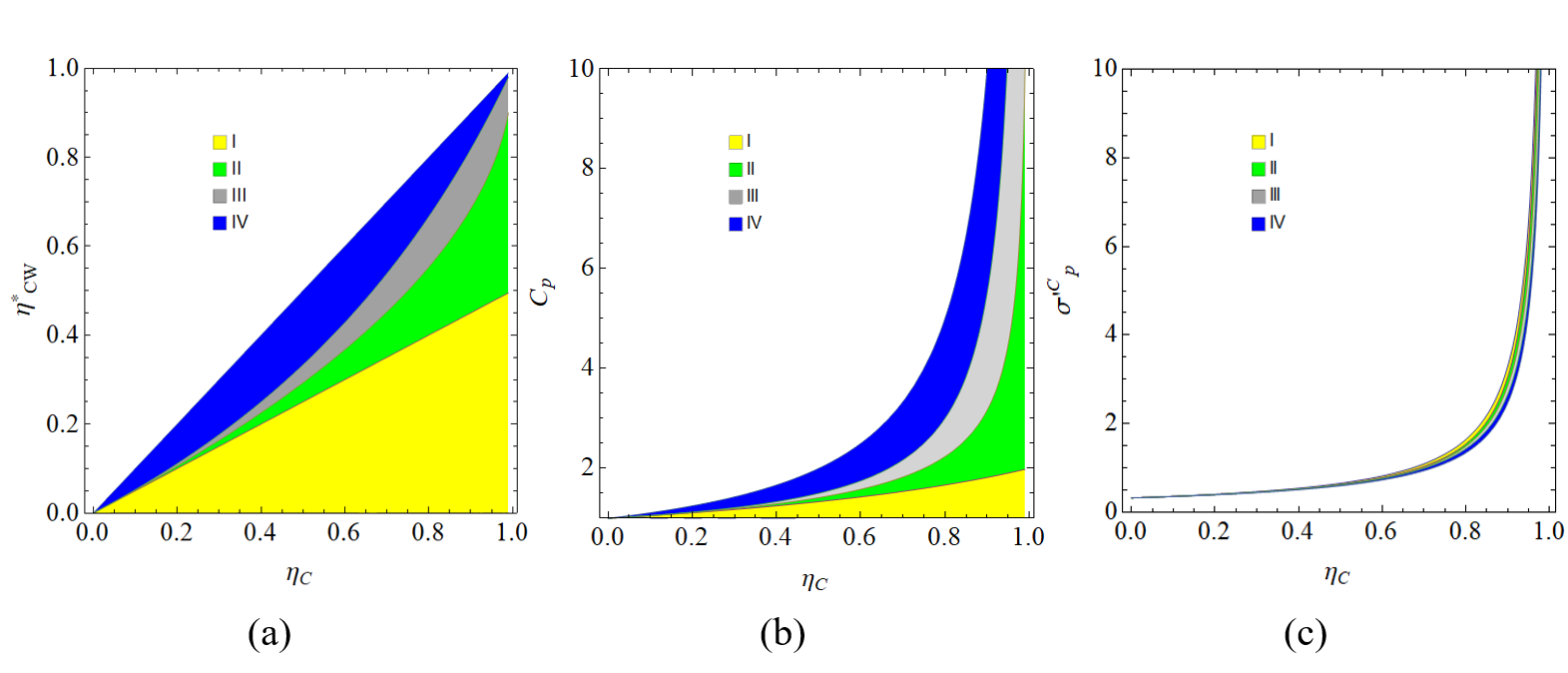}
	\caption{(a) $2D$ mapping of the efficiency at maximum power $\eta^{*}_{CW}$ in Eq. (\ref{eff2}) vs Carnot efficiency $\eta_{C} = 1 - \tau$ . 
	(b) 2D mapping of the $c_p$ vs $\eta_C$ corresponding to (a). (c) 2D mapping of the $\sigma'^{C}_p$ vs $\eta_C$ corresponding to (a)} \label{et1}
\end{figure*}
Entire parameter space corresponding to the efficiency given by Eq. (\ref{eff2}) can be separated in four regions summarized in Table I represented by the colorful 2D shapes in Fig. \ref{et1}a. Using the dimensionless pump frequency $c_p$ as a control parameter which depends on the effective temperature ratio $\tau$, the pump pulse bandwidth $\sigma'^{C}_{p}$ (the classical pump bandwidth of $\sigma'_{p}$) that depends on the dimensionless probe coupling field $\lambda'$, $\tau$ and $u, v$ and $\alpha$. We define the characteristic efficiency values describing the boundaries between the four regions correspond to $0$, $\eta_C/2$ (between I and II regions), $\eta_C/(2-\eta_C)$ (between III and IV), Carnot efficiency $\eta_C=1-\tau$ upper bound of IV) and Curzon-Ahlborn (CA) limit \cite{klr14} $\eta_{CA}=1-\tau^{1/2}$ (between II and III).  Note, that the two parameters of the pump field: the frequency $\omega_p$ and the Rabi frequency $\Omega_p$ which define an effective hot bath temperature $T_h$ and the pump bandwidth $\sigma'^{C}_{p}$ can be controlled experimentally. Thus the 2D parameter space $\{\tau,c_p\}$ and $\{\tau, \sigma'^{C}_{p} \}$ shows a constrained relation between the two as seen in Figs. \ref{et1}b and \ref{et1}c, respectively. 

We now compare the two-classical-photon pump with our previous work \cite{qd21} where a single resonant pump has been taken to drive transition $g-2$. Let us highlight some important points here. First, the range of the pump frequency, in the single photon is $\omega_p \ge 2\omega_c$ while in the two photon case $\omega_p \ge \omega_c$, which affirms that the size of the system can be smaller. Second, the particular boundary is reached at different pump parameters. For instance, a CA limit is obtained for the single photon pump at $c_p \simeq 2/\sqrt{\tau}$ and for the two photons $c_p \simeq 1/\sqrt{\tau}$ and its corresponding Rabi frequency is $\Omega^{CA}_{p} \simeq (4\omega_{p}/\omega_{c})^{4} T^{2}_{c}\delta/ \Gamma^{2}_{2}$. The factor of two, which appears in the other bounds as well, originates from the quadratic scaling of the photon absorption probability with the input intensity for the classical light.

\subsection{Entangled-two-photon pump}
We now consider the case when the two photons driving transitions $g \rightarrow e$ and $e \rightarrow 2$ are entangled. A classical pump beam at frequency $w_p$ directed into a crystal is down converted into entangled photon pair - signal $(s)$ and idler $(i)$ with frequencies $\omega_s$ and $\omega_i$ , respectively, as shown in Fig. 5a. We consider type-II down conversion, which corresponds to orthogonally polarized signal and idler beams which allows us to introduce the entanglement time and avoid complications with the selection rules. The different group velocities along two polarization axes create a time delay between the signal and the idler photons represented by entanglement time $T$. Owing to energy conservation, $\omega_{p} = \omega_{i} + \omega_{s}$. The photon pair is fully characterized by the twin photon state amplitude $\varphi (\omega_{i}, \omega_{s}) = \mathcal{A}(\omega_i + \omega_s) \Phi(\omega_i, \omega_s)$, where $\mathcal{A}(\omega) = \frac{A_{0}}{\omega - \omega_{p} + i\sigma}$ is a Lorentzian envelop function of the pump photon with bandwidth $\sigma$ centered around $\omega_{p}$ and $\Phi(\omega_i, \omega_s) = \text{sinc} \left[(\omega_{s} - \omega_i)T/2 \right] $, originating from the phase matching inside the crystal (see Appendix B).
\begin{figure}
	\includegraphics[width=0.47\textwidth]{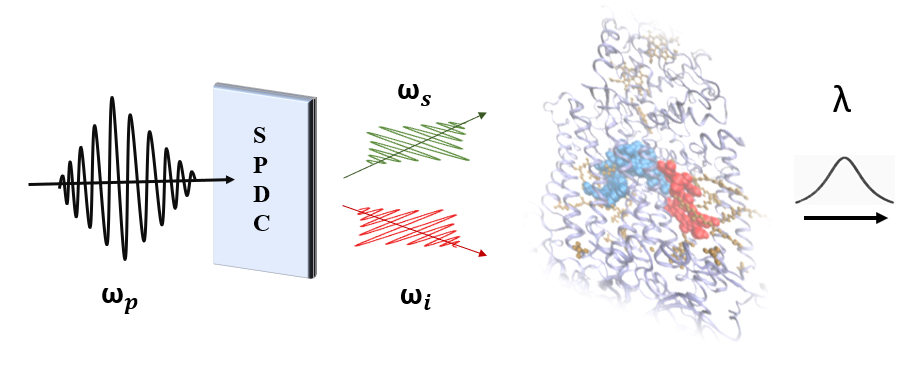}
	\caption{(a) The pump photon frequency $\omega_p$ is down converted into two signal and idler photons with frequencies $\omega_s$ and $\omega_i$, respectively. We consider spontaneous down conversion (SPDC) with energy conservation $\omega_{p} =\omega_{i} + \omega_{s}$. The entangled photons drive transitions $g \rightarrow e'$ and $e' \rightarrow 2$ and the probe field then stimulates the emission from	state $1$ to the excited vibrational level $0$ of the ground electronic state.} \label{qfig}
\end{figure}
The leading contribution to population of state $2$ calculated perturbatively according to the ladder diagram in Fig. \ref{feyd} and by following the same approach given in Appendix A to obtained Eq. (\ref{eq14}) and assuming that transition energy $\omega_{21}$ is much larger than dephasing rate $\Gamma_2$, the final population reads
\begin{align}
\varrho_{11}(t) &= \frac{\mathbb{N}^{2} \, {\Gamma_2} \,\tilde{n}_2 \,\omega_{2e'} \,\omega_{e'g} \,\tilde{\omega}^{2}_{2g} \left( 1 - e^{-\Gamma _2 \left(2 n_2 + 1 \right) t} \right) \theta} { \Gamma _2 \left(2 n_2 + 1 \right) \left( \sigma^{2}_{p} + \tilde{\omega}^{2}_{2g} \right)^2}, \nonumber\\
\varrho_{gg}(t) &= 1 - \varrho_{11}(t) \label{p1q}
\end{align}
where $\tilde{n}_{2} = n_{2}+1, \,\tilde{\omega_{2g}} = \omega_{2g} - \omega_{p},  \theta = \text{sinc}^{2} [ \frac{T (\omega_{2e'} - \omega_{e'g})}{2} ]\ $, and $\mathbb{N}^{2} = \frac{\mathcal{N}^{2} A_0^2\mu_{eg} \mu_{e'e} \mu_{2e'}\mu_{21} }{4 \epsilon^{2}_{0} V^{2}}$ is a normalization. Assume for brevity, that the normalization  $\mathbb{N}$ of quantum state is same as that of a classical \cite{fenfmuk2021}, $\mathbb{N} = \Omega_{1'} \Omega_{2'}$. This ensures that all the dimensionless parameters for the entangled case are the same as in the classical case. Similarly to the classical case we now introduce an effective hot bath characterized by the thermal photon occupation number $n_h$ and dephasing rate $\Gamma_h$ which drives the transition $g \leftrightarrow 1$, where the parameters of the bath are given by 
\begin{align}
n_h&=\frac{\tilde{n}_2 \,\theta \, \omega _{2 e'} \omega _{g e'} \,\Omega^{2}_{1'} \Omega^2_{2'} \Delta^2 } {\left(2 n_2+1\right) \tilde{\Delta}^4 - 2 \Omega^{2}_{1'} \Omega^2_{2'} \Delta^2 \tilde{n}_2 \omega _{2 e'} \omega _{g e'} \theta} \nonumber\\
\Gamma _h&=\frac{\Gamma _2 \big(\left(2 n_2+1\right) \tilde{\Delta}^4 - 2 \,\tilde{n}_{2} \, \theta \,\omega _{2 e'} \omega _{g e'} \Omega^{2}_{1'} \Omega^2_{2'} \Delta^2 \big)} {\tilde{\Delta}^4},\label{ppmq} \ \ \ \ \ \
\end{align}
where $\tilde{\Delta}^{2} = \Delta^2 + \sigma^2_{p}$, and $\Delta = \omega_{2g} - \omega_{p}$.
\begin{figure*}
	\includegraphics[width=0.9\textwidth]{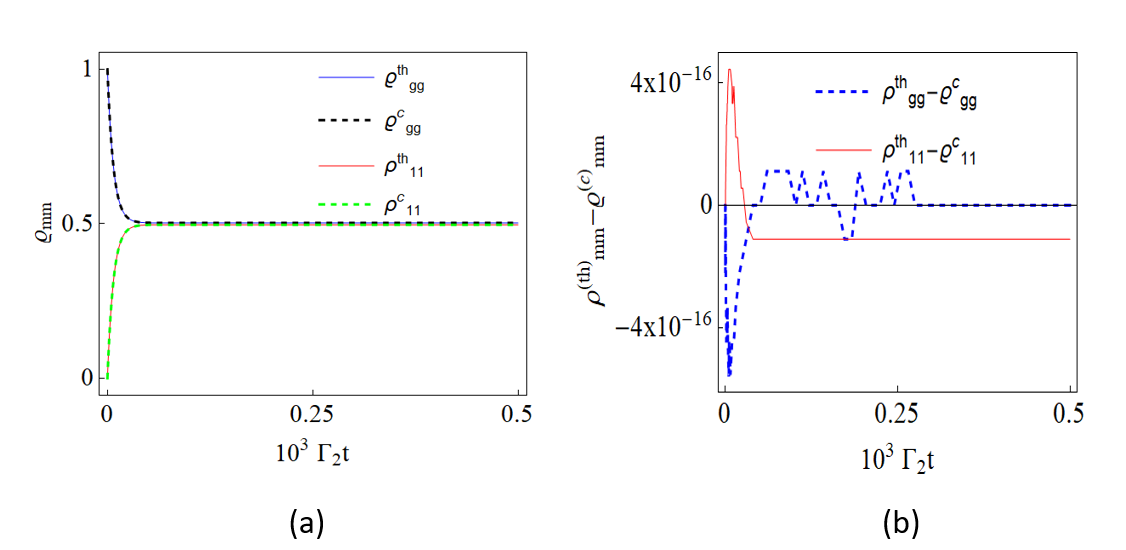}
	\caption{(a)The population of ground $g$ and lowest excited state $1$ obtained using a coherent bath in Eq. (\ref{p1q}) and a thermal bath $\rho_{th}$ using parameters in Eq. (\ref{ppmq}). (b) The difference between populations of coherent and thermal baths $\varrho_c - \rho_{th}$ for parameters in Eq. (\ref{ppmq}). }\label{qpm}
\end{figure*}
Using the effective bath introduced for the system excited by the two entangled photons in Eq. (\ref{ppmq}), we can perfectly match the populations of $g$ and $1$ driven by the thermal bath given in Eq. (\ref{teq2}) as shown in Fig. 6. In the two-photon entangled, pump bandwidth scale: $\sigma'^{Q}_{p} = \sigma^{e}_{p} \Gamma_{2}/T_{c} \Delta$, where $\sigma^{e}_{p} = (\sigma^{2}_{p} - \Delta^{2}) ^{1/2}$ 

\begin{figure*}
	\includegraphics[width=0.95\textwidth]{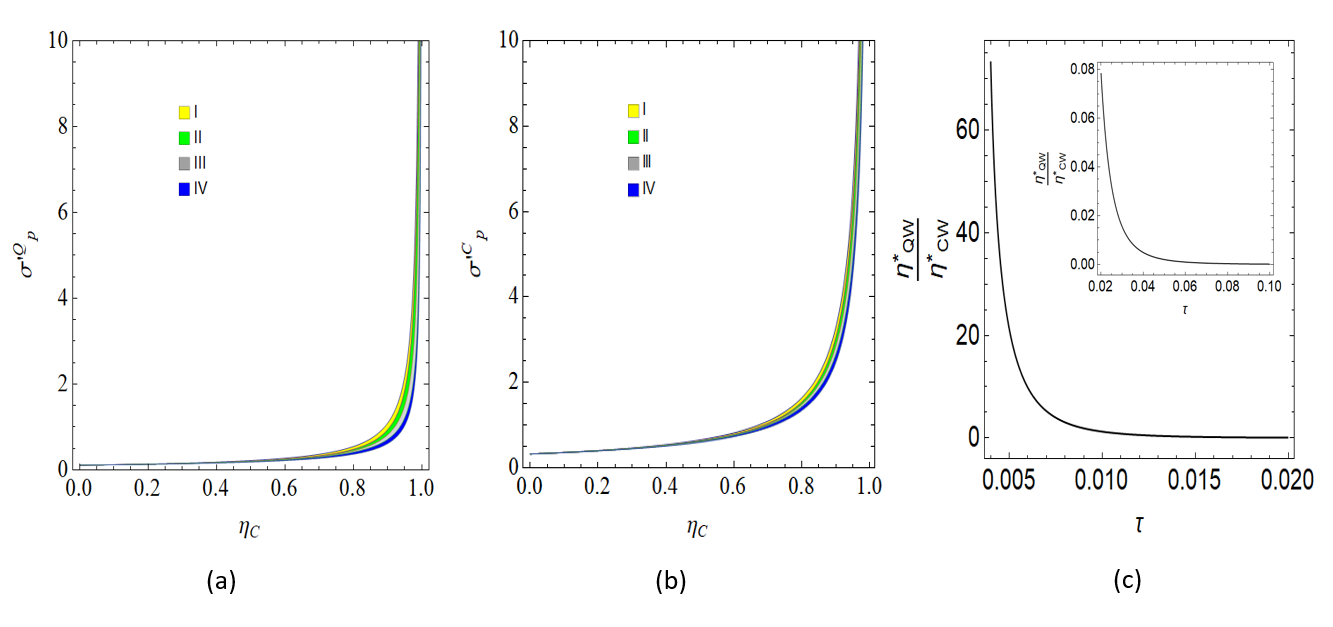}
	\caption{ (a) The 2D mapping of $\sigma'^{Q}_{p} \text{vs}\  \eta_{C}$ for entangled photons  given in Table. 2. (b) The 2D mapping of $\sigma'^{C}_{p} \text{vs} \ \eta_{C}$ of two pump photons from Table 1. (c) The ratio of efficiency at maximum power in Eqs. (\ref{qeff2}) and (\ref{eff2}) vs $\tau$ for $\text{sinc}^{2} \left[\frac{T(\omega_{2e'} - \omega_{ge'})}{2}\right] \sim 1$ and $\eta^{*}_{QW} > \eta^{*}_{CW}$ occurs for small $\tau$ also in inset $\eta^{*}_{QW} < \eta^{*}_{CW}$ for considerably large value of $\tau$. The parameters read: $T_{2} = T_{c} = 300 \text{K}, \omega_{p} = 1.3 \text{ev}, \omega_{c} = 0.012 \text{ev}, \Omega_{p}= 0.023 \text{ev}, \lambda = 0.1 \text{ev}, \delta = 0.00003 \text{ev}, \sigma_{p} = 200 \text{cm}^{-1}, \Gamma_{2} = 0.71 \text{ps}^{-1}$ and $\Gamma_{c} = 0.025 \text{ps}^{-1} $. } \label{cvq}
\end{figure*}

Following the general approach outlined in Ref. \cite{qd21} we apply the high temperature limit for the phonon bath $i.e.,T_2 >> \omega_{21}$, and the maximum power w.r.t $c_{21}$ which is recast in terms of dimensionless parameters given in Eq. (\ref{pq26}) yields:
\begin{align}\label{qpmx}
P^{max}_{Q} = \frac{4  u v \lambda' \EuScript{W}  \tau ^4 c'^{2}_p (\sigma'^Q_p)^{4} \left(\theta - \tau ^4  \sigma'^4_p\right)}{3 \theta \left(\EuScript{X} + \tau ^4 v \ c'_p  (\sigma'^Q_p)^{4} \right) \left(\EuScript{X} +\tau ^4 u c'_p \lambda' (\sigma'^Q_p)^{4} \right)},
\end{align}
where $\EuScript{X} = \EuScript{W}+ \EuScript{E}, \EuScript{E} = \alpha \,  u \, v \, \text{sinc}^{2}[ T \left(\omega _{2 e'}-\omega _{g e'}\right)/2] $, $\EuScript{W} = \sqrt{u v \left(\tau ^4 c'_p \sigma'^4_p + \EuScript{E}/v \right) \left(\tau ^4 c'_p \lambda ' \sigma'^4_p + \EuScript{E}/u \right)}$, and $c'_p = c_p - 1$. The corresponding efficiency at maximum power is given by
\begin{align}
\eta^{*}_{Q} = 1-\frac{1}{c_p -\frac{ (c_p-1) \EuScript{E} }{\EuScript{W} + \EuScript{E}}}\label{qeff}, 
\end{align}
where subscript $Q$ denotes the two photon entangled pump. Similarly, in the weak  dissipating regime Eq. (\ref{qeff}) can be recast as
\begin{align}
\eta^{*}_{QW} = 1-\frac{1}{c_p + \frac{ u \, v \,\alpha ^2 \, \theta^{2}}{\tau ^4 \lambda' \sigma'^4_p \left(\alpha \, u \,\theta + \left(c_p - 1\right) \tau ^4 \sigma'^4_p  \right)}}, \label{qeff2}
\end{align}
where subscript $QW$ specifies the weak dissipating limit of two-photon entangled (quantum) pump and $\theta = \text{sinc}^{2}[T \left(\omega _{2 e'} - \omega _{g e'}\right)/2 ]$. Similar to the classical efficiency given in Eq. (\ref{eff2}) the entire parameter space of the respective quantum efficiency in Eq. (\ref{qeff2}) is also divided into four regions summarized in Table 2. By comparing Tables 1 and 2, it is clear that the four regions for $\eta^{*}_{CW}$  and $\eta^{*}_{QW}$ are identical when $\sigma'^{C}_{p} = \sigma^{'Q}_{p}$. The distinction between $\sigma^{'C}_p$  and $\sigma^{'Q}_{p}$ originate due to additional parameter $T$ and the different pump intensity scaling (quadratic vs linear) \cite{dsm16} as mentioned in Tables 1 and 2. The effective bandwidth for the two-classical and two-entangled photon pump vs $\eta_{C}$ is depicted in Figs. \ref{cvq}a and \ref{cvq}b, respectively. Furthermore, the efficiency corresponding to the maximum output power for the quantum light is more robust than that for the classical light for the moderate range of $\tau$ as shown in Fig. \ref{cvq}c and which will be discussed in the next subsection. 
\begin{table}
\begin{tabular}{c c c c}
\hline\hline
	Bound &$\eta^{*}_{QW}$& $c_p$ & $\sigma'^{Q}_{p}$ \\[0.2cm] \hline
	I &0&$1$& $\sqrt[4]{\Xi \left(u - \sqrt{\frac{u \left(u \lambda '-4 v\right)}{\lambda '}} \right) }$\\[0.2cm] 
	I/II&$\frac{\eta_{C}}{2}$ &$\frac{2}{2-\eta_{C}}$ & $\sqrt[4]{\Xi \left(u - \sqrt{ \frac{u \left(\lambda' u-2 v \left(2-\eta _C\right)\right)}{\lambda '}} \right)}$ \\[0.2cm] 
	II/III &$\eta_{CA}$&$\frac{1}{\sqrt{1-\eta_{C}}}$ & $\sqrt[4]{\Xi \left(u - \sqrt{\frac{u \left(u \lambda '-4 v \sqrt{1-\eta _C}\right)}{\lambda '}}\right)}$ \\[0.2cm] 
	III/IV &$\frac{\eta_{C}}{2-\eta_{C}}$&$\frac{2-\eta_{C}}{2(1-\eta_{C})}$& $\sqrt[4]{\Xi \left(u - \sqrt{\frac{u \left(u \left(2-\eta _C\right) \lambda '-8 v \left(1-\eta _C\right)\right)}{\left(2-\eta _C\right) \lambda '}}\right)}$ \\[0.2cm] 
	IV  &$\eta_{C}$&$\frac{2}{1-\eta_{C}}$& $\sqrt[4]{\Xi \left(u - \sqrt{\frac{u \left(u \lambda '-4 v \left(1-\eta _C\right)\right)}{\lambda '}}\right)}$ \\[0.2cm] \hline
	\hline \label{tab2}
	\end{tabular}
	\caption{The efficiency and pump scale are same as in Table I. The pump bandwidth parameters of the quantum bath corresponding to the QHE efficiency bounds in this table shown in Fig. \ref{cvq}, where $\Xi = \frac{ \alpha}{2 \left(1-\eta _C\right){}^4}  \text{sinc}^{2} [ \frac{T(\omega_{2e'} - \omega_{ge'})}{2} ]$.}
\end{table}

\subsection{Maximum QHE power for the quantum and the classical two-photon pump}
After maximization of QHE power with respect to the temperature and pump bandwidth we obtain different scaling for the classical and the entangled two photon pumps \cite{sdm18}. Fig. \ref{ncq}a shows numerically that in the specific temperature range the maximum output power in entangled case can be larger than that in the classical case. The quantum enhancement for the maximum output power occurs for small $\tau$. In this case Eqs. (\ref{eq21}) and (\ref{qpmx}) yield, respectively,
\begin{align}
P^{max}_{C} &=&\frac{\tau ^8 \left(c_p-1\right){}^2 \lambda' (\sigma'^c_p)^{8}}{3 \,\alpha } \label{cp}
\end{align}
\begin{align}
 P^{max}_{Q} &=& \frac{\tau ^4 \left(c_p-1\right){}^2 \lambda' (\sigma'^Q_p)^{4}} {3 \, \alpha \, \text{sinc}^{2} \left[ \frac{T(\omega_{2e'} - \omega_{ge'})}{2} \right]}, \label{cq}
\end{align}
where for brevity, $\sigma'^{C}_{p} = \sigma'^{Q}_{p} = \sigma'_{p}$, and from Eqs. (\ref{cp}) and (\ref{cq}) results, we have $P^{max}_Q / P^{max}_C = \tau^{-4} \sigma'^{-4}_p \text{sinc}^{-2} [ T(\omega_{2e'} - \omega_{ge'})/2 ] $ which gives  $P^{max}_Q > P^{max}_C$ for $\text{sinc}^{2} [ T(\omega_{2e'} - \omega_{ge'})/2 ] \simeq 1$ and $\sigma'_{p} \tau < 1$. The above analysis clearly indicates the relation between the effective bath temperature, the entanglement time and the spectral bandwidth of the optical fields as well as the system energy scale and its effect on the optical measurements with the entangled light in the open quantum systems. For instance in the limit of short entanglement time we can achieve quantum enhancement even in highly anharmonic system as long as $|\omega_{2e'} - \omega_{e'g}|<<1/T$. Similarly, for the long entanglement time the quantum enhancement can be reached for nearly harmonic system $(\omega_{2e'} \simeq \omega_{e'g})$. In the same time inequality $\sigma'_{p}\tau < 1$ yields an additional requirement for the pumping source such that $\Omega_{p}> 4 \delta (\sigma^{e}_{p})^{2}/\Delta^{2}$.
\section{Spectroscopic regime}
So far we have discussed QHE regime in which the density matrix  has been solved nonperturbatively. We now focus on the pump-probe spectroscopic signal derived by perturbative approach in light - matter interaction.  Ref. \cite{qd21} shows the apparent connection between the thermodynamics of the QHE and the spectroscopy which emerges as an incoherent control tool for the optimization of optical measurements, which can enhance the yield of fluorescence, pump-probe measurements, and improve the signal-to-noise ratio in a wide class of the optical signals. Here we explore the class of two-photon pump - classical probe signals using classical and entangled two-photon pumps. To that end the coherence $\rho_{01}$ and $\rho_{10}$ which enters in the definition of Eq. (\ref{eq:Pdef}) can be calculated perturbatively. 

Using Eq. (\ref{A1}) (in Appendix A), while we keeping the leading order terms following the Feynman diagram in Fig. 2, we substitute the solution for the population $\rho_{11}(t)$ from Eq. (\ref{eq14}) and solve for  $\rho_{01}(t)$ for two-photons pump, which yields
\begin{align}
	\rho_{01} &= -\frac{32 \, i \, \Gamma _2 \,\lambda \, n_2 \,\delta ^4 \, \Omega ^4{}_p}{\left(\delta ^2+4 \sigma^2_{p} \right)^4 \sigma _{\text{pr}} \left(\Gamma _c n_c + \Gamma _2 n_2\right) \left(5 \Gamma _2 n_2 + \Gamma_c n_c \right)},\nonumber\\ \rho_{10} &= -\rho_{01}.\label{c33} 
\end{align}
Similarly, we can obtained the coherence $\varrho_{01}(t)$ while we keeping the leading order terms following the Feynman diagram in Fig. 2, we substitute the solution for the population $\varrho_{11}(t)$ from Eq. (\ref{p1q}) and we get
\begin{align}
	\varrho_{01} &= -\frac{2 \, i \, \Gamma _2 \,\lambda \, n_2 \,\delta ^4 \, \Omega ^2{}_{1'}\Omega ^2{}_{2'} \text{Sinc}^{2}  \left[ \frac{T(\omega_{2 e'} - \omega_{g e'})}{2}  \right]}{\left(\Delta ^2 + \sigma^2_{p} \right)^4 \sigma _{\text{pr}} \left(\Gamma _c n_c + \Gamma _2 n_2\right) \left(5 \Gamma _2 n_2 + \Gamma_c n_c \right)},\nonumber\\ \varrho_{10} &= -\varrho_{01}.\label{qc33} 
\end{align}
Utilizing the Eq. (\ref{c33}) we obtained the power for the classical two-photon pump in Appendix C, Eq. (\ref{sp34}). After optimizing w.r.t $c_{21}$, the maximum power yields
\begin{align}
\mathcal{P}^{max}_{C} = \frac{u \lambda ' \left( 3 c_p+5 \alpha  u - 3 - \mathcal{C} \right)}{2 \tau ^8 \sigma _{\text{pr}} \sigma'^8_p},\label{cmp36} \ 
\end{align}
where $\mathcal{C} = \sqrt{5 \left(c_p+\alpha  u-1\right) \left(c_p+5 \alpha  u-1\right)}$. Similarly, we optimize the power for the entangled two-photon pump in Eq. (\ref{qp35}) and it maximum yields
\begin{align}
\mathcal{P}^{max}_{Q} = \frac{4 \alpha  c_{21} u^2 \left(c_p-c_{21}-1\right) \lambda' \text{sinc}^{2} \big[\frac{T \left(\omega _{2 e'}-\omega _{g e'}\right)}{2} \big]}{\tau ^4 \sigma _{\text{pr}} \left(c_{21}+\alpha  u\right) \left(c_{21}+5 \alpha  u\right) \sigma'^4{}_p}. \label{qmp37}
\end{align}
In appendix C, we demonstrated numerically that the maximum power for the quantum light is much larger than that for the classical light within for the moderate range of $\tau<1$ in Eqs. (\ref{qmp37}) and (\ref{cmp36}) (see Fig. \ref{ncq}b). The ratio of maximum power for the corresponding equations reads
\begin{align}
\frac{\mathcal{P}^{max}_{Q} }{\mathcal{P}^{max}_{C}} = \tau ^4 \sigma'^4_p \text{sinc}^{2} \big[\frac{T \left(\omega _{2 e'}-\omega _{g e'}\right)}{2} \big]. \label{eq23}
\end{align}
For the short entanglement time $\text{sinc}^{2} \big[\frac{T \left(\omega _{2 e'}-\omega _{g e'}\right)}{2} \big] \simeq 1$ and in the limit of $\sigma'_{p} \tau > 1$ we obtain $\mathcal{P}^{max}_{Q} > \mathcal{P}^{max}_{C}$. We therefore identified the parameter regime where maximum power for the entangled two-photon pump is enhanced compared to the classical case using perturbative regime. In comparison to QHE (nonperturbative) regime the power increase due to the entanglement in spectroscopic (perturbative)  regime occurs when the bath temperature ratio  is $\tau>1/\sigma_p'$, whereas in the former case $\tau<1/\sigma_p'$, which agrees with strong pumping (nonperturbative) vs weak pump (perturbative) limit taken in these two cases.

The simulation of the maximum power for QHE regime given in Eqs. (\ref{cp}) and (\ref{cq}), and for Spectroscopic regime given in Eqs. (\ref{cmp36}), and (\ref{qmp37}), respectively shown in Fig. 8 (Appendix C). It shows that the quantum enhancement for the power is achieved at different time scale: in the QHE regime at $0 < \tau <0.5 \text{X}10^{-2}$, and in spectroscopic regime at $0 < \tau <0.07$.
\section{Summary}
It has been shown that the two photon absorption of entangled light may enhance the Raman excitation \cite{sac21} due to different intensity scaling at low photon fluxes.  In the present analysis the two photon absorption in open quantum system regime benefits from additional control parameters using incoherent control scheme by mimicking the QHEs. In this proposed model we analytically explored the characteristics of two photon absorptions for the classical and entangled pair of photons and their dependence on additional degrees of freedom, due to which we get the maximum work, in both for weak and strong intensity approximations.  By using the approach of \cite{qd21} we developed connection between the thermodynamics of the QHE and the spectroscopy. The transfer of entanglement to the system allows to optimize the detailed balance in system-bath driven optical transitions in an open quantum system allowing QHE to operate near thermodynamic cycle, which consequently provides an enhanced yield of conversion between the pump and the probe fields. Our results can be further extended to  Raman, hyper-Raman and other techniques that require additional control over illumination intensity and pump light statistics.

\section{ACKNOWLEDGMENTS}
We gratefully acknowledge the support from the National Science Foundation of China (No. 11934011), the Zijiang Endowed Young Scholar Fund, the East China Normal University and the Overseas Expertise Introduction Project for Discipline Innovation (111 Project, B12024). M.Q. acknowledges the support from CSC Scholarship (CSC No. 2018 DFH 007778).
\appendix

\setcounter{equation}{0}
\renewcommand{\theequation}{A\arabic{equation}}
\begin{center}
	\textbf{\large{Appendix A: Effective heat bath }}
\end{center}
The master equation in Liouville space
	\begin{equation}
		\frac{d\hat{\rho}}{dt} =  - \frac{i}{\hbar} \left[ \hat{H}_{int}(t),~ \hat{\rho}\right], \label{meq}
	\end{equation} 
	We define the superoperator in Liouville space by acting on a arbitrary operator $X$
	\begin{equation}
		\hat{H}_{int, -} \hat{X} = \hat{H}_{int} \hat{X} - \hat{X} \hat{H}_{int}. \label{so}
	\end{equation} 
	Now, the solution of Eq. (\ref{meq}) can be written as a Dyson series in Liouville space. Hence, we obtain the density matrix
	\begin{equation}
		\hat{\rho}(t) = \mathcal{G}(t,t_0)\hat{\rho}(t_0), \label{rt}
	\end{equation}
	with the Liouville space Green’s function
	\begin{equation}
		\mathcal{G}(t,t_0) = \hat{\mathcal{T}} \left[ -\frac{i}{\hbar}\text{exp}\int_{t_0}^{t} \hat{H}_{int, -}(\tau) \right]
	\end{equation}
	where $\mathcal{T}$ is a time ordering superoperator which is defined by
	\begin{align}
		\mathcal{T}A(t_1)B(t_2) &\equiv \Theta(t_1 - t_2) A(t_1)B(t_2) + \Theta(t_2 - t_1)\nonumber \\ &B(t_2) A(t_1),
	\end{align}
	where $A(t)$ and $B(t)$ are two arbitrary superoperators and $\Theta (t)$ is the Heaviside function. A perturbative expansion of the Dyson series yields the leading order contribution of Eq. (\ref{rt}) reads off:
	\begin{align}
		\rho_{11}(t) &= \frac{1}{\hbar^{4}} \int_{t_0}^{t}d\tau_{1}\int_{t_0}^{\tau_{1}} d\tau_{2} \int_{t_0}^{t_2} d\tau_{3} \int_{t_0}^{\tau_3} d\tau_{4}<< H_{int,-}\nonumber\\ &(\tau_1) H_{int,-}(\tau_2) H_{int,-}(\tau_3) H_{int,-}(\tau_4) \rho({t_{0})}>>,
	\end{align}
	where $<<\cdot>>=\text{Tr}\left[\cdot,\rho(t) \right]$ represents the trace with the density operator. The population of the excited state due to relaxation of $2-1$ is represented by a diagrammatic Feynman ladder in Fig. 2 given by
\begin{align}
		\rho^{a}_{11}(t) & = \frac{1}{\hbar^{4}} \int^{t} d\tau_{1} \int^{\tau_{1}} d\tau_{2} \int^{\tau_{2}} d\tau_{3} \int^{\tau_{3}} d\tau_{4} \langle \langle \mathcal{G}_{11,22} \nonumber\\ & (t - \tau_1) \hat{\mathcal{V}}_{2e} \mathcal{G}_{2e,2e}(\tau_1 - \tau_2) \hat{\mathcal{V}} ^{\dagger}_{2e} \mathcal{G}_{ee,e'e'}(\tau_2 - \tau_3)  \hat{\mathcal{V}}_{e'g} \nonumber\\ & \mathcal{G}_{e'g,e'g}(\tau_3 - \tau_4) \hat{\mathcal{V}} ^{\dagger}_{e'g} \rangle \rangle \langle \hat{E}^{\dagger}_{1}(\tau_3) \hat{E}^{\dagger}_{2}(\tau_1) \hat{E}_{2}(\tau_2) \hat{E}_{1}(\tau_4) \rangle, \nonumber\\
		&=\frac{1}{\hbar^{4}} \int^{t} dt_{1} \int^{t_{1}} dt_{2} \int^{t_{2}} dt_{3} \int^{t_{3}} dt_{4} \langle \langle \mathcal{G}_{11,22}(t_1)\nonumber\\ &\hat{\mathcal{V}}_{2e} \mathcal{G}_{2e,2e}(t_2) \hat{\mathcal{V}} ^{\dagger}_{2e} \mathcal{G}_{ee,e'e'}(t_3)  \hat{\mathcal{V}}_{e'g}  \mathcal{G}_{e'g,e'g}(t_4) \hat{\mathcal{V}} ^{\dagger}_{e'g} \rangle \rangle \nonumber\\ & \langle \hat{E}^{\dagger}_{1}(t-t_{1}-t_{2}-t_{3}) \hat{E}^{\dagger}_{2}(t-t_1) \hat{E}_{2}(t-t_{1}-t_{2}) \nonumber\\ &\hat{E}_{1}(t-t_{1}-t_{2}-t_{3}-t_{4}) \rangle,
\end{align}
where $t_1 = t - \tau_{1}, t_2 = \tau_1 - \tau_2, t_3 = \tau_2 - \tau_{3}, t_4 = \tau_3 - \tau_{4}$ and $\mathcal{V}$ is a time independent dipole operator in Liouville space. Now, we transform the fields components in frequency domain and it can be recast as
\begin{widetext}
		\begin{align}
			\rho^{a}_{11}(t) &= \frac{1}{\hbar^{4}} \int \frac{d\omega_{1} \  d\omega_{1'} \ d\omega_{2} \ d\omega_{2'}}{(2\pi)^{4}} \langle \hat{\Omega}^{\dagger}_{1}(\omega_{1'}) \hat{\Omega} ^{\dagger} _{2}(\omega_{2'}) \hat{\Omega}_{2}(\omega_{2}) \hat{\Omega}_{1} (\omega_{1})\rangle \int_{-\infty}^{\infty} dt_{1} dt_{2} dt_{3} dt_{4}\mathcal{G}_{11,22}(t_1) \mathcal{G}_{2e,2e}(t_2) \nonumber\\
			& \mathcal{G}_{ee,e'e'} (t_3)\mathcal{G}_{e'g,e'g}(t_4) e^{i \omega_{1'}(t - t_1 - t_2 - t_3) + i \omega_{2'}(t - t_1) - i\omega_{2} (t - t_1 - t_2) - i\omega_{1}(t - t_1 - t_2 - t_3 - t_4) },\nonumber\\
			&= \frac{1}{\hbar^{4}} \int \frac{d\omega_{1} \  d\omega_{1'} \ d\omega_{2} \ d\omega_{2'}}{(2\pi)^{4}} \langle \hat{\Omega}^{\dagger}_{1}(\omega_{1'}) \hat{\Omega} ^{\dagger}_{2'}(\omega_{2'}) \hat{\Omega}_{2}(\omega_{2}) \hat{\Omega}_{1}(\omega_{1}) \rangle e^{i(\omega_{1'} + \omega_{2'} - \omega_{1} - \omega_{2})t}\nonumber\\
			&\mathcal{G}_{11,22}(\omega_{1} + \omega_{2} - \omega_{2'} - \omega_{1'}) \mathcal{G}_{2e,2e}(\omega_{1} + \omega_{2} - \omega_{1'}) \mathcal{G}_{ee,e'e'} (\omega_{1} - \omega_{1'}) \mathcal{G}_{e'g,e'g}(\omega_{1}), \label{fey14}\ \ \ \
		\end{align}
\end{widetext}
		where $\hat{\Omega}_{1}(\omega_1) = \mu_{ge}\sqrt{\frac{\hbar {\omega}_1}{2 \mathbf{V}\epsilon_{0}}} \hat{a}_{1}(\omega_1)e^{-i\omega_{1}t}$, similarly we define $\hat{\Omega}_{2} (\omega_2), \hat{\Omega}^{\dagger}_{1}(\omega_1')$ and $\hat{\Omega} ^{\dagger}_{2} (\omega_2' )$.
	
The population Green's functions $\mathcal{G}_{11,22}, \mathcal{G}_{ee,e'e'}$, and $\mathcal {G}_{gg,00} $ originate from the solution of coupled transport (relaxation) equations:
	\begin{align}
		\dot{\rho}_{22} &=\nonumber -\Gamma_{2} (n_{2}+1)\rho_{22} + \Gamma_{2} n_{2} \rho_{11},\\
		\dot{\rho}_{11} &= \Gamma_{2} (n_{2}+1)\rho_{22} - \Gamma_{2} n_{2} \rho_{11},
		\label{eq.1}
	\end{align}
	\begin{align}
		\dot{\rho}_{e'e'} &=\nonumber -\Gamma_{e} (n_{e}+1)\rho_{e'e'} + \Gamma_{e} n_{e} \rho_{ee},\\
		\dot{\rho}_{ee} &= \Gamma_{e} (n_{e} + 1)\rho_{e'e'} - \Gamma_{e} n_{e} \rho_{ee},
		\label{eq.e}
	\end{align}
	\begin{align}
		\dot{\rho}_{00} &= \nonumber-\Gamma_{c} (n_{c}+1)\rho_{00} + \Gamma_{c} n_{c} \rho_{gg},\\
		\dot{\rho}_{gg} &= \Gamma_{c} (n_{c}+1)\rho_{00} - \Gamma_{c} n_{c} \rho_{gg}.
		\label{eq.1a}
	\end{align}
	Eqs. (\ref{eq.1}) - (\ref{eq.1a}) can be recast as a Pauli master equation:
	\begin{align}
		\dot{\rho}_{ii}(t) &= -\sum_{ii,jj}^{}\kappa_{ii,jj} \rho_{jj}(t),
		\label{eq.2p}
	\end{align}
	where, $\kappa_{ii,jj}$ is the population transport matrix.  In Eq. (\ref{eq.2p}), the diagonal elements, $i=j$, $\kappa_{ii,ii}$ are positive, whereas the off-diagonal elements, $i \neq j$, $\kappa_{ii,jj}$ are negative. The population transport matrix satisfies the population conservation:   $\sum_{i}^{}\kappa_{ii,jj}=0$. The evolution of the diagonal elements is defined by the population Green function, $\rho_{jj}(t) = \sum_{i}\mathcal{G}_{jj,ii}(t)\rho_{ii}(0)$. where $\mathcal{G}_{jj,ii}(t)$ is given \cite{mcr9}
	\begin{align}
		\mathcal{G}_{jj,ii}(t) = \sum_{n}^{}\xi_{jn}^{(R)} D_{nn}^{-1}\exp(-\lambda_{n}t) \xi_{ni}^{(L)},
		\label{eq.2g}
	\end{align}
	where $\lambda_{n}$ is the $nth$ eigenvalue of left and right eigenvector $(\xi_{n}^{(L)},\xi_{n}^{(R)})$ and $D = \xi^{L} \xi_{R}$ is a diagonal matrix. Using Eq. (\ref{eq.2g}) we obtain for the population Green's functions:
	\begin{align}
		\mathcal{G}_{00,gg}(t) &= \frac{n_c(1- e^{-t(1+2 n_{c})\Gamma_{c} })  }{(1+2  n_{c})}\label{g0},\\
		\mathcal{G}_{ee,e'e'}(t) &= \frac{n_e(1- e^{-t(1+2 n_{e})\Gamma_{e} })  }{(1+2  n_{e})}\label{ge},\\
		\mathcal{G}_{11,22}(t) &= \frac{(1+n_2)(1- e^{-t(1+2 n_{2})\Gamma_{2} })  }{(1+2  n_{2})}.
		\label{g12}
	\end{align}
Utilizing Eqs. (A14 - \ref{g12}) and Liouville space Green's functions $\mathcal{G}(\omega) = \frac{-(n_i + 1)\Gamma_{i}}{(\omega + i\epsilon) \left[\omega + i(2 n_i + 1)\Gamma_{i} \right]}$, where $n_i $ is the average phonon occupation number and $\Gamma_i$ is the dephasing rate for the $i \leftrightarrow i-1$ transition. To examine field-induced fourth order correlations of matter, we utilize the reduced density matrix obtained by tracing out the field degrees of freedom, Eq. (\ref{fey14}) reads
\begin{widetext}
	\begin{align}
		\rho^{a}_{11}(t) &= \frac{\Gamma_{2} \Gamma_{e} (n_2 + 1)(n_e + 1) e^{-(\Gamma_{21} + \epsilon_{2})t}} {\hbar^{4}(\epsilon_{2} - \Gamma_{21}) (\epsilon_{e} - \Gamma_{ee'})} (e^{\Gamma_{21}t}  \hat{\Omega} ^{\dagger}_{2} [\ \omega_{2e} + i(\epsilon_{2} - \Gamma_{2e})]  - e^{\epsilon_{2}t} \hat{\Omega}^{\dagger}_{2} [\ \omega_{2e} + i(\Gamma_{21} - \Gamma_{2e}) ] ) \nonumber\\ & \hat{\Omega}_{1}\left[\omega_{e'g} - i\Gamma_{ee'}\right] \big(  \hat{\Omega}_{2} [\ \omega_{2e} + i(\epsilon_{e} - \Gamma_{2e} )] \hat{\Omega} ^{\dagger}_{1}[\ \omega_{e'g} + i(\epsilon_{e} - \Gamma_{e'g}) ] - \hat{\Omega}_{2} [\ \omega_{2e} - i(\Gamma_{2e} - \Gamma_{ee'})] \nonumber\\ & \hat{\Omega}^{\dagger} _{1} [\ \omega_{e'g} + i(\Gamma_{ee'} - \Gamma_{e'g})] \big),\label{eqf1} 
\end{align}
where $\epsilon_2 ~\text{and}~ \epsilon_{e}$ are the dephasing rate at transition $2-1$ and $e-e'$, respectively.

The population of vibrational state $1$ from Feynman diagram b and c: 
\begin{eqnarray}
		\rho^{b}_{11}(t) &=& \frac{1}{\hbar^{4}} \int^{t} dt_{1} \int^{t_{1}} dt_{2} \int^{t_{2}} dt_{3} \int^{t_{3}} dt_{4} \langle \langle \mathcal{G}_{11,22}(t-t_1) \hat{\mathcal{V}}^{\dagger}_{2e} \mathcal{G}_{e2,e2} (t_1 - t_2)  \hat{\mathcal{V}}_{2e} \mathcal{G}_{ee,e'e'}(t_2 - t_3) \nonumber\\ && \hat{\mathcal{V}}_{e'g} \mathcal{G}_{e'g,e'g}(t_3 - t_4) \hat{\mathcal{V}} ^{\dagger}_{eg} \rangle \rangle \langle \hat{E}^{\dagger}_{1}(t_3) \hat{E}^{\dagger}_{2}(t_2) \hat{E}_{2}(t_1) \hat{E}_{1}(t_4) \rangle \nonumber \\
		&=& \frac{\Gamma_{2} \Gamma_{e} (n_2+1)(n_e+1) e^{-(\Gamma_{21} + \epsilon_{2})t}} {\hbar^{4}(\epsilon_{2} - \Gamma_{21}) (\epsilon_{e} - \Gamma_{ee'})} (e^{\Gamma_{21}t}  \hat{\Omega}_{2}\left[ \omega_{2e} - i(\epsilon_{2} - \Gamma_{2e}) \right]  - e^{\epsilon_{2}t} \hat{\Omega}_{2} \left[ \omega_{2e} - i(\Gamma_{21} - \Gamma_{2e}) \right] ) \hat{\Omega}_{1}\left[\omega_{e'g} - i\Gamma_{e'g}\right] \nonumber\\ && \big( \hat{\Omega}^{\dagger}_{1} \left[\omega_{e'g} + i(\epsilon_{e} - \Gamma_{e'g} )\right] \hat{\Omega}^{\dagger}_{2}\left[\omega_{2e} - i(\epsilon_{e} - \Gamma_{2e}) \right] - \hat{\Omega}^{\dagger}_{1} \left[\omega_{e'g} - i(\Gamma_{e'g} - \Gamma_{ee'})\right] \hat{\Omega}^{\dagger}_{2}\left[\omega_{2e} - i(\Gamma_{ee'} - \Gamma_{2e})\right] \big).\label{eqf2} \\
		\rho^{c}_{11}(t) &=& \frac{1}{\hbar^{4}} \int^{t} dt_{1} \int^{t_{1}} dt_{2} \int^{t_{2}} dt_{3} \int^{t_{3}} dt_{4} \langle \langle \mathcal{G}_{11,22}(t-t_1) \hat{\mathcal{V}}_{2e'} \mathcal{G}_{2e',2e'} (t_1 - t_2)  \hat{\mathcal{V}}_{e'g} \mathcal{G}_{2g,2g}(t_2 - t_3) \nonumber\\ && \hat{\mathcal{V}}^{\dagger}_{2e'} \mathcal{G}_{e'g,e'g}(t_3 - t_4) \hat{\mathcal{V}}^{\dagger}_{eg} \rangle \rangle \langle \hat{E}^{\dagger}_{1}(t_2) \hat{E}^{\dagger}_{2}(t_1) \hat{E}_{2}(t_3) \hat{E}_{1}(t_4) \rangle\nonumber \\
		&=& \frac{\Gamma_{e} (n_e+1) e^{-(\Gamma_{21} + \epsilon_{2})t}} {\hbar^{4}(\Gamma_{21}-\epsilon_{2})} (e^{\Gamma_{21}t}  \hat{\Omega}^{\dagger}_{2}\left[ \omega_{2e'} + i(\epsilon_{2} - \Gamma_{2e'}) \right]  - e^{\epsilon_{2}t} \hat{\Omega}^{\dagger}_{2} \left[ \omega_{2e'} + i(\Gamma_{21} - \Gamma_{2e'}) \right] )  \hat{\Omega}_{1}\left[\omega_{e'g} - i\Gamma_{e'g}\right] \nonumber\\ && \hat{\Omega}_{2} \left[\omega_{2g} - \omega_{e'g} + i(\Gamma_{e'g} - \Gamma_{2e} )\right] \hat{\Omega}^{\dagger}_{1}\left[ \omega_{2g} - \omega_{2e'} + i(\Gamma_{2e'} - \Gamma_{2g}) \right]. \label{eqf3} 
	\end{eqnarray}
\end{widetext}

The The total population $\rho_{11}(t) = \text{Re} (\rho^{a}_{11} + \rho^{b}_{11} + \rho^{c}_{11})$ is induced by pulse with Lorentzian shape is defined $e.g.$ $\hat{\Omega}_{1} \left[\omega_{e'g} + i \Gamma_{e'g} \right] = \frac{\hat{\Omega}_{1'}} {\omega_{e'g}-\omega_{0} + i\sigma_{p} - i\Gamma_{e'g}},$ where $\omega_{0} $ is the central frequency and $\omega_{e'g}$ is the $e'-g$ transition frequency. Employing the approximation $\Gamma_{21} = \Gamma_{2}(2 n_2 + 1) >> \epsilon_{2}$, assuming that the transition energy is much larger than the decay constant. The total population of level $1$ by employing Eqs. (\ref{eqf1}), (\ref{eqf2}) and (\ref{eqf3}) yields: 

\begin{align}
	\rho_{11}(t) = \frac{ 16 (n_2+1) \, \Omega^{4}_{p} \tilde{\omega}^{2}_{2e'} \, \tilde{\omega}^{2}_{e'g} (1 - e^{\Gamma_{2}(2 n_2 + 1)t}) }{ (2 n_2 + 1) (\sigma^{2}_{p} + \tilde{\omega}^{2}_{2e'} )^{2} (\sigma^{2}_{p} + \tilde{\omega}^{2}_{e'g} )^{2}},\label{A22}
\end{align}
where $\tilde{\omega}_{2e'} = \omega_{2e'} - \omega_{0}, \tilde{\omega}_{e'g} = \omega_{e'g} - \omega_{0}$ and we set $|\Omega_{1'}| = |\Omega_{2'}| = |\Omega^{\dagger}_{1'}| = |\Omega^{\dagger}_{2'}| = |\Omega_p|$.

We consider a three-level molecular system with the ground state $g$, single electronic excited state $e$ and double excited electronic state $f$ shown in Fig. 1(a). we denote the viberational states of electronic ground states $0$ and $g$ and vibrational double excited states of electronic states $2$ and $1$ and $e$ and $e'$ for single electronic states.
	The corresponding equation of motion for the density matrix is given by
	\begin{widetext}
		\begin{align}
			\dot{\rho}_{gg}(t) &= i\Omega_{1} ( \rho_{ge}(t) - \rho_{eg}(t) ) + \Gamma_{c}(n_{c} + 1) \rho_{00}(t) - \Gamma_{c}n_{c}\rho_{gg}(t),\nonumber\\
			\dot{\rho}_{ee}(t) &= -i\Omega_{1}(\rho_{ge}(t) - \rho_{eg}(t)) + i\Omega_{2}(\rho_{g2}(t) - \rho_{2e}(t)),\nonumber\\
			\dot{\rho}_{22}(t) &= -i\Omega_{2}(t)(\rho_{e2}(t) - \rho_{2e}(t)) - \Gamma_{2}(n_{2} + 1)\rho_{22}(t)+\Gamma_{2}n_{2}\rho_{11}(t),\nonumber\\
			\dot{\rho}_{11}(t) &=-i \lambda (\rho_{01}(t) - \rho_{10}(t)) - \Gamma_{2}(n_{2} + 1)\rho_{00} (t) + \Gamma_{c}n_{c} \rho_{gg}(t), \nonumber\\
			\dot{\rho}_{2e}(t) &= i\Omega_{2}(t)(\rho_{22}(t) - \rho_{ee}(t)) + i\Omega_{1} \rho_{2g}(t) - \left\{ \frac{\Gamma_{2}(n_{2}+1)}{2} + i(\omega_{2e} - \nu_{2})\rho_{2e} \right\}, \nonumber\\
			\dot{\rho}_{eg}(t) &= i\Omega_{1}(\rho_{ee}(t) - \rho_{gg}(t)) - i\Omega_{1}\rho_{2g}(t) - \left\{\frac{\Gamma_{c}n_{c}}{2} + i(\omega_{eg} - \nu_{1}) \right\}\rho_{eg}(t), \nonumber\\
			\dot{\rho}_{2g}(t) &= i\Omega_{1}\rho_{2e}(t) - i\Omega_{2}\rho_{eg}(t) - \left\{ \frac{\Gamma_{2}(n_2 + 1)}{2} + \frac{\Gamma_{c}n_{c}}{2} + i(\omega_{2g} - \nu_{1} - \nu_{2}) \right\}\rho_{2g}(t), \nonumber\\
			\dot{\rho}_{10}(t) &= i\lambda (\rho_{11}(t) - \rho_{00}(t)) - \left\{ \frac{\Gamma_{2}n_{2}}{2} + \frac{\Gamma_{c}(n_{c} + 1)}{2} + i(\omega_{10} - \nu_{pr}) \right\} \rho_{10}(t),\nonumber\\
			\dot{\rho}_{e2}(t) &= \dot{\rho}^{\dagger}_{2e},~ \dot{\rho}_{ge}(t) = \dot{\rho} ^{\dagger}_{eg}, ~\dot{\rho}_{g2}(t) = \dot{\rho}^{\dagger}_{2g},~ \dot{\rho}_{01}(t) = \dot{\rho}^{\dagger}_{10},\label{A1}
		\end{align} 
	\end{widetext}
where $\Gamma_{2}/2$ is a dephasing rate and $n_{2} = \left[ \text{exp}(\hbar \omega_{21}/ k_{B} T_{c} )\right]$ is the average phonon occupation number corresponding to $1 \leftrightarrow 2$ at temperature $T_{2}$.\\
The output power in Eq. (\ref{genp}) using Eq. (\ref{ppm}) in the high temperature limit becomes: 
\begin{align}
	P_{C} \!= \!\frac{4\, T_{c}\, \omega_{c}\, c_{21}\, \tau ^8 u \, v \, \tilde{c}_{p} \lambda ' \sigma '^{8}_p \left( \tilde{c}_{21} - \tau ^8 \left(\tilde{c}_{21} - c_{21}\right) \sigma'^{8}_p \right)} {3 \, (c_{21}\, \tau ^8 \, \sigma'^{8}_p + u \, \tilde{c}_{21} ) \left(c_{21} \tau ^8 \lambda ' \sigma'^{8}_p + v \tilde{c}_{21} \right)}, \ \ \ \ \label{eq20}
\end{align}
where subscript $C$ specifies the two photons pump and $\tilde{c}_{p} = \left(c_p - c_{21} - 1\right), \tilde{c}_{21} = \alpha - c_{21}, u = \Gamma_{2} \omega_{c}/ \Gamma_{c} T_{c}$, $v = \Gamma_{c}/ \omega_{c}$ and $\alpha = T_{2}/ \omega_{c}$, where $T_2$ is the phonon bath temperature of level $1-2$. 

\setcounter{equation}{0}
\renewcommand{\theequation}{B\arabic{equation}}
\begin{center}
	\textbf{\large{Appendix B: Entangled states of two photon}}
\end{center}
The state of SPDC to first order in perturbation theory: 
\begin{align}
|\psi \rangle = |0 \rangle - \frac{i}{\hbar}\int_{t_{0}}^{t} \text{dt} \  \mathcal{H}_ {I}(t)|0 \rangle,\label{eq1}
\end{align}
where $\mathcal{H}_{I}$ is the effective third-order interaction Hamiltonian given by
\begin{align}
\mathcal{H}_{I}(t) = \epsilon_{0} \int_{V} \text{d}^{3} \boldsymbol{r}  \chi^{(2)} E^{+}_{p}(\boldsymbol{r},t) \hat{E}^{-}_{s} (\boldsymbol{r},t) \hat{E}^{-}_{i} (\boldsymbol{r},t) + \text{h.c.},\label{2}
\end{align} 
where $\chi^{(2)}$ is the susceptibility tensor of rank $2$ which describe the nonlinear crystal. $V$ is interaction volume covered by pump filed and the pump field is simply chosen a classical plane wave along z direction.
\begin{align}
E^{+}_{p}(z,t) = E_{p} \int \text{d}\omega_{p}  \mathcal{A} (\omega_{p}) e^{ -i(k_{p}(\omega_{p})z - \omega_{p}t)},\label{3}
\end{align}
where $\mathcal{A}(\omega_{p})$ is the pulse envelope function.. 
\begin{align}
\hat{E}^{-}_{j}(z,t) = \int \text{d}\omega_{j} \mathcal{E}(\omega_{j}) \hat{a}^{\dagger}_{j}(\omega_j) e^{-i(k_{j}(\omega_j) z - \omega_j t)},\label{4}
\end{align}
where $j=s,i$ and $\hat{a}^{\dagger}(\omega_{j})$ is a creation operator and we have restricted the spatial integral to be over z coordinate only. We assume that $\mathcal{E}(\omega_j) = \sqrt{\hbar \omega/\epsilon_{0} \textbf{V} }$, where $\textbf{V}$ is quantization volume is slowly varying over the frequencies of interest and therefore we can bring it outside the integral. The interaction Hamiltonian part of Eq. (\ref{eq1}) using Eqs. (\ref{2}), (\ref{3}) and (\ref{4}) is recast as
\begin{align}
&\int_{t_{0}}^{t} \mathcal{H}_{I} (t') \text{d}t' = \mathcal{A} \int_{-\infty} ^{+\infty} \text{d}t' \int_{\frac{-L}{2}}^{\frac{+L}{2}} \text{d}z \int \text{d}\omega_{i} \text{d}\omega_{s}\text{d} \omega_{p} \nonumber\\ & e^{-i(k_{i}(\omega_{i}) + k_{s}(\omega_{s}) - k_{p}(\omega_{p}))z} e^{i(\omega_{s} + \omega_{i} - \omega_{p})t} \mathcal{A}(\omega_{p}) \nonumber\\ & \hat{a}^{\dagger}_{i} (\omega_i)\hat{a}^{\dagger}_{s}(\omega_s) + \text{h.c.},\label{5}
\end{align}
where $L$ is the length of the crystal and $\mathcal{A} = E_{p} \mathcal{E}(\omega_i) \mathcal{E}(\omega_s)$. For a pulsed laser, we can
assume that the pump field, and therefore the interaction Hamiltonian, is zero before $t_{0}$ and after $t$. Therefore, we can extend the limits of the integration over infinite time. Performing the time integral yields $\delta(\omega_i + \omega_s - \omega_p)$, which then allows the $\omega_p$ integral to be evaluated, giving
\begin{align}
	&\int_{t_{0}}^{t} \mathcal{H}_{I} (t') \text{d}t' = -2 \pi \mathcal{A} \int_{\frac{-L}{2}}^{\frac{+L}{2}} \text{d}z \int \text{d}\omega_{i} \text{d}\omega_{s} \mathcal{A}(\omega_{i}+\omega_{s}) \nonumber\\ &  e^{-i(k_{i}(\omega_{i}) + k_{s}(\omega_{s}) - k_{p}\omega_{p})z} \hat{a}^{\dagger}_{i}(\omega_i)\hat{a}^{\dagger}_{s}(\omega_s) + \text{c.c.}, \ \ \label{6}
\end{align}
Evaluating the integral over z yields and substitute in Eq. (\ref{eq1})
\begin{align}
	|\psi\rangle  &= |0\rangle + \frac{2 i\pi L \mathcal{A}}{\hbar} \int \text{d}\omega_{i} \text{d}\omega_{s}    \mathcal{A}(\omega_{i} + \omega_{s}) \Phi(\omega_{i}, \omega_s) \nonumber \\& \hat{a}^{\dagger}_{i}(\omega_i)\hat{a}^{\dagger}_{s}(\omega_s) |0\rangle  + \text{H.C.},\label{7}
\end{align}
where $\Phi (\omega_{i}, \omega_s) = \text{sinc}(\frac{L\Delta{k}}{2})$ and $\mathcal{A}(\omega_{s},\omega_{i}) = \frac{A_{0}}{\omega_{i} + \omega_{s} - \omega_{p} + i\sigma}$ is the normalized band pump pulse of width $\sigma $. For narrow band pump $\sigma \rightarrow 0$. 

The essential character of the phase-matching function is better illustrated when it is expressed in a simpler form obtained
by making the Taylor expansions 
\begin{align}
k_{p}(\omega) &\approx& k_{p0} + (\omega - \bar{\omega})\frac{\partial k_{p}(\omega)}{\partial \omega}|_{\omega=2\bar{\omega}} ,\nonumber \\ k_{j}(\omega) &\approx& k_{j0} + (\omega - \bar{\omega}) \frac{\partial k_{j}(\omega)}{\partial \omega}| _{\omega=\bar{\omega}} \label{8}
\end{align}
Here, $2 \bar{\omega}$ is the center pump frequency.
Discarding all but the first two terms yields by using Eq. (\ref{8})
\begin{align}
\Delta k &= k_{s}(\omega_s) + k_{i}(\omega_i) - k_{p}(\omega_s + \omega_i)\nonumber \\ &\approx  (\omega_{s} - \omega_{i})( \frac{\partial k_{s} (\omega_s)}{\partial  \omega_{s}} - \frac{\partial k_{i} (\omega_i)}{\partial  \omega_{i}})\nonumber\\
&= (\omega_{s} - \omega_{i})(\frac{1}{v_{s}} - \frac{1}{v_{i}}). \label{9}
\end{align}
Therefore, the phase matching factor reads:
\begin{align}
\Phi(\omega_{s},\omega_{i}) = \text{sinc}(\frac{(\omega_{s} - \omega_{i})T}{2}),\label{10}
\end{align}
where, $T = L(\frac{1}{v_{s}} - \frac{1}{v_{i}})$  is entanglement time, characterizing the group velocity dispersion inside the SPDC crystal. The output state of SPDC from Eq. (\ref{eq1}) is given by
\begin{align}
|\psi \rangle = \mathcal{N} A_{0} \int\int \frac{\text{d}\omega_{s}\text{d} \omega_{i} \Phi(\omega_s, \omega_{i})}{\omega_i + \omega_s - \omega_p + i\sigma}  \hat{a}^{\dagger}_{s}(\omega_{s}) \hat{a}^{\dagger}_{i}(\omega_{i}) |0\rangle, \label{12}
\end{align}
where $\mathcal{N}$ is normalization constant. 

\begin{figure*}
	\includegraphics[width=0.99\textwidth]{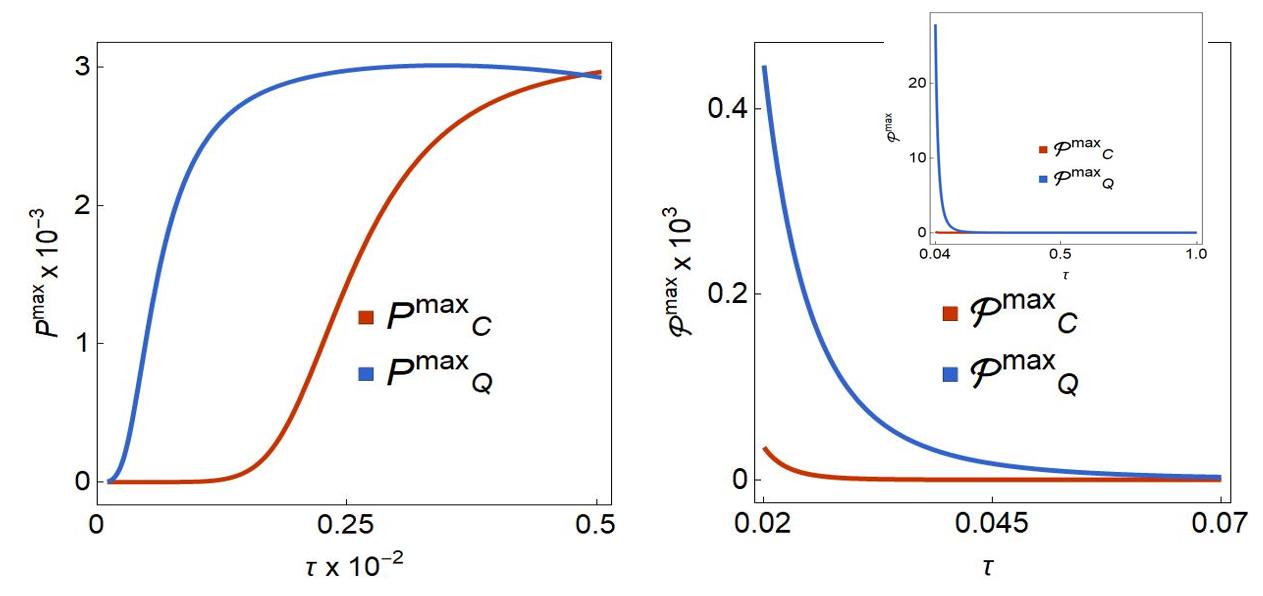}
	\caption{ (a) The numerical simulation of maximum classical power vs $\tau$ and it shows the within the small range of $\tau $ shows quantum advantage. (b) same as (a) for the spectroscopic power shows the quantum advantage for the different regime of temperature scale. The vertical axis of (a) and (b) correspond to the maximum classical and quantum output power of the non-perturbed and perturbed regimes, respectively and in this simulation, the quantum advantages differ in their magnitude as shown in Figs. The parameters read: $T_{2} = T_{c} = 300 \text{K}, \omega_{p} = 1.3 \text{ev}, \omega_{c} = 0.012 \text{ev}, \Omega_{p}= 0.023 \text{ev}, \lambda = 0.1 \text{ev}, \delta = 0.00003 \text{ev}, \sigma_{p} = 200 \text{cm}^{-1}, \Gamma_{2} = 0.71 \text{ps}^{-1}$ and $\Gamma_{c} = 0.025 \text{ps}^{-1} $ }\label{ncq}
\end{figure*}

For the classical light 
\begin{align}
\langle \hat{E}^{\dagger}_{1}(\omega_1)\hat{E}^{\dagger}_{2}(\omega_2) \hat{E}_{2}(\omega_2) \hat{E}_{1}(\omega_1) \rangle &= |\langle 0 | \hat{E}_{2}(\omega_2) \hat{E}_{1} (\omega_1) |\phi\rangle |^{2}\nonumber\\
&= |\hat{E}_{2}(\omega_2)\hat{E}_{1}(\omega_1)|^{2}\label{13}
\end{align}
The Rabi frequency for the transition from $g-2$ via intermediate level $e$ in a given classical field 
\begin{align}
\Omega^{2}_1 \ \Omega^{2}_2 &=& |\mu_{eg}|^{2}|\mu_{2e}|^{2} |E_{2}(\omega_2)E_{1}(\omega_1)|^{2} \nonumber\\
&=& |\mu_{eg}|^{2}|\mu_{2e}|^{2} |\langle 0 | E_{2}(\omega_2)E_{1} (\omega_1) \phi\rangle|^{2}. \label{14} \ \ \ \
\end{align}
For quantum light, the two-point field correlation function reads 
\begin{align}
&\langle 0| \hat{E}_{2} (\omega_2) \hat{E}_{1} (\omega_1) | \psi\rangle = \langle 0| \sqrt{\frac{\omega_{s}}{2 \epsilon_{0} V}} \hat{a}_{2}(\omega_{s}) \sqrt{\frac{\omega_{i}}{2 \epsilon_{0} V}} \hat{a}_{1}(\omega_{i}) |\psi\rangle \nonumber\\
&=\frac{\mathcal{N}}{2\epsilon_{0} V}  \int\int \text{d} \omega_{s} \text{d} \omega_{i} \sqrt{\omega_{s} \omega_{i}} \Phi(\omega_{s}, \omega_{i}) \frac{A_{0}}{\omega_{i}+\omega_{s} - \omega_{p} + i\sigma }\nonumber \\
&  \underline{\langle 0| \hat{a}_{2}(\omega_{s})\hat{a}_{1}(\omega_{i}) \hat{a}^{\dagger}_{i}(\omega_{i}) \hat{a}^{\dagger}_{s}(\omega_{s}) |0\rangle } \rightarrow \delta_{2s} \delta_{1i}\nonumber\\
&=\frac{A_{0}\mathcal{N}}{2\epsilon_{0} V} \frac{\sqrt{\omega_1 \omega_2}}{\omega_{1} + \omega_{2} - \omega_{p} + i\sigma } \Phi( \omega_2, \omega_1). \! \label{15}
\end{align}
By combination of Eqs. (\ref{10}), (\ref{12}), (\ref{14}) and (\ref{15})
the quantum and correlation functions recast as
\begin{align}
{\Omega}_1(\omega_1) {\Omega}_2(\omega_2) &= \frac{\mu_{eg}\mu_{2e}\mathcal{N} A_{0}  }{2\epsilon_{0} V} \frac{\sqrt{\omega_2 \omega_1}}{\omega_{1} + \omega_{2}-\omega_{p} + i\sigma } \nonumber\\ & \text{sinc} [\frac{(\omega_2 - \omega_1)T}{2} ]
\end{align}
The output power in Eq. (\ref{genp}) using Eq. (\ref{ppmq}) in the high temperature limit becomes: 
\begin{align}
	P_{Q} = \frac{2 \, u \,  v \, c_{21}  \tilde{c}_p \lambda' \,\tau ^4  \sigma'^4_p \left(2 \, \theta \, \tilde{c}_{21} - \tau ^4 \left(2 \tilde{c}_{21} - c_{21}\right) \sigma'^4_p\right)}{3 \left(u \ \tilde{c}_{21} \theta + c_{21} \tau^4 \sigma'^4_p\right) \left(v \ \tilde{c}_{21} \theta + c_{21} \lambda' \tau ^4 \sigma'^4_p \right)}, \ \ \label{pq26}
\end{align}
where subscript $Q$ denotes the two photon entangled pump and $\tilde{c}_{21} = \alpha+c_{21}, \tilde{c}_{p} = c_{p} - c_{21} -1, \theta = \text{sinc}^{2} [T \left(\omega _{2 e'}-\omega _{g e'}\right)/2], \alpha = T2/\omega_{c}, u = \Gamma_{2} \omega_c/\Gamma_{c}T_{c}$ and $v = \Gamma_{c}/\omega_{c}$.
\setcounter{equation}{0}
\renewcommand{\theequation}{C\arabic{equation}}
\begin{center}
	\textbf{\large{Appendix C: Maximum power of QHEs and Spectroscopy }}
\end{center}
Using Eq. (\ref{c33}) in the definition of the power of Eq. (\ref{eq:Pdef}) and applying dimensionless parameters over the high temperature limit, the spectroscopic power is given by
\begin{align}
		\mathcal{P}_{C} = \frac{4 \alpha  c_{21} u^2 \left(c_p-c_{21}-1\right) \lambda '}{\tau ^8 \sigma _{\text{pr}} \left(c_{21}+\alpha  u\right) \left(c_{21}+5 \alpha  u\right) \sigma'^8_p}, \label{sp34}
	\end{align}
	where all dimensionless parameters defined in earlier sections. Similarly we recast spectroscopic power for entangled two-photon source using Eq. (\ref{qc33})
	\begin{align}
		\mathcal{P}_{Q} = \frac{\alpha  c_{21} u^2 \left(c_p-c_{21}-1\right) \lambda ' \text{sinc}^{2} \big[\frac{T \left(\omega _{2 e'}-\omega _{g e'}\right)}{2} \big]}{2 \tau ^2 \sigma _{\text{pr}} \left(c_{21}+\alpha  u\right) \left(c_{21}+5 \alpha  u\right) \sigma'^2_p} \label{qp35}.
\end{align}
The maximum output power given in Eq. (\ref{eq21}) and Eq. (\ref{qpmx}) for two-photon entangled and classical states respectively and their numerical simulation vs $\tau$ shown in Fig. \ref{ncq}(a), we considered small interval of $\tau $ because only in this regime it shows quantum advantages in the scale of $10^{-3}$ and within the range of $\tau \in \left[0, 0.0048 \right]$ the maximum output power of two-photon entangled pump larger than the two photon classical pump case. The small value of $\tau$ correspond to the high intensity of the pump field, because $\tau = T_{c}/T_{h}$, where $T_{h} \propto \sqrt{\Omega_{p}}$ and $T_{c}$ at room temperature.
Similarly, the maximum output power for spectroscopic regime in Eqs. (\ref{cmp36}) and (\ref{qmp37})  shown in Fig. \ref{ncq}(b) and the quantum advantage shown in the regime of low pump intensity. The analytical solution of maximum output power in the limit of $\tau$ explained in main text.


\begin{thebibliography}{9}
\bibitem{c24}
S. Carnot, \textit{R\'eflections sur la Puissance Motrice du Feu et sur les Machines propres \`a D\'evelopper cette Puissance}. Paris: Bachelier, (1824).

\bibitem{sb59}
H. E. D. Scovil, E. O. Schulz-DuBois, Three-Level Masers as Heat Engines. Phys. Rev. Lett. 2, 262 (1959).

\bibitem{sbj59}
J. E. Geusic, E. O. Schulz-DuBois, and R. W. De Grasse, and H. E. D. Scovil, Three level spin refrigeration and maser action at 1500 mc/sec, J. Appl. Phys. 30, 1113 (1959).

\bibitem{sb67}
J. Geusic, E. O. Schulz-Du Bois, H. E. D. Scovil, Quantum Equivalent of the Carnot Cycle. Phys. Rev. 156, 343 (1967).

\bibitem{bh13}
Brandao, F. G. S. L., Horodecki, M., Oppenheim, J., Renes, J. M. and Spekkens, R. W. Resource theory of quantum states out of thermal equilibrium. Phys. Rev. Lett. 111, 250404 (2013).

\bibitem{bm17}
Bera, ManabendraNath, Riera, A., Lewenstein, M. and Winter, A. Generalized laws of thermodynamics in the presence of correlations. Nat. Commun. 8, 2180 (2017).

\bibitem{bc19}
F. Binder, L. A. Correa, C. Gogolin, J. Anders, and G. Adesso (eds.), Thermodynamics in the Quantum Regime (Springer,	Cham, Switzerland, 2019).

\bibitem{rq84}
R. Kosloff, A quantum mechanical open system as a model of a heat engine, J. Chem. Phys. 80, 1625 (1984).

\bibitem{jmc16}
J. Jaramillo, M. Beau, and A. del Campo, Quantum supremacy of many-particle thermal machines New J. Phys. 18,	075019 (2016).

\bibitem{kbn21}
W.K. Mok, K. Bharti, L.C. Kwek, and A. Bayat, Optimal probes for global quantum thermometry, Communications	physics 4, 1 (2021).

\bibitem{sga21}
Sushant Saryal, Matthew Gerry, Ilia Khait, Dvira Segal, and Bijay Kumar Agarwalla, Universal Bounds on Fluctuations in Continuous Thermal Machines. arXiv:2103.13513v1.

\bibitem{bg17}
Thomas G, Banik M, Ghosh S. Implications of Coupling in Quantum Thermodynamic Machines. Entropy. 2017; 19(9):442. https://doi.org/10.3390/e19090442.

\bibitem{ud21}
Jaegon Um, Konstantin E. Dorfman, and Hyunggyu Park. Coherence enhanced quantum-dot heat engine. arXiv:2111.09582v1

\bibitem{kmp14}
J. V. Koski, V. F. Maisi, J. P. Pekola, and D. V. Averin, Experimental realization of a Szilard engine with a single electron,	Proc. Natl. Acad. Sci. USA 111, 13786 (2014).

\bibitem{jdt16}
J. Roßnagel, S. T. Dawkins, K. N. Tolazzi, O. Abah, E. Lutz, F. Schmidt-Kaler, and K. Singer, A single-atom heat engine, Science 352, 325 (2016).

\bibitem{kf17}
J. Klaers, S. Faelt, A. Imamoglu, and E. Togan, Squeezed Thermal Reservoirs as a Resource for a Nanomechanical Engine	Beyond the Carnot Limit, Phys. Rev. X 7, 031044 (2017).

\bibitem{pb19}
J. P. S. Peterson, T. B. Batalhão, M. Herrera, A. M. Souza,	R. S. Sarthour, I. S. Oliveira, and R. M. Serra, Experimental Characterization of a Spin Quantum Heat Engine, Phys. Rev.
Lett. 123, 240601 (2019).

\bibitem{lgs19}
D. von Lindenfels, O. Gräb, C. T. Schmiegelow, V. Kaushal, J. Schulz, M. T. Mitchison, J. Goold, F. Schmidt-Kaler, and U. G. Poschinger, Spin Heat Engine Coupled to a Harmonic Oscillator Flywheel, Phys. Rev. Lett. 123, 080602 (2019).

\bibitem{kbl19}
J. Klatzow, J. N. Becker, P. M. Ledingham, C. Weinzetl, K. T. Kaczmarek, D. J. Saunders, J. Nunn, I. A. Walmsley, R. Uzdin, and E. Poem, Experimental Demonstration of Quantum Effects
in the Operation of Microscopic Heat Engines, Phys. Rev. Lett. 122, 110601 (2019).

\bibitem{nba21}
Bouton, Q. Nettersheim, J. Burgardt, S. Adam, D. Lutz, E. Widera, A. A quantum heat engine driven by atomic collisions. Nat.	Commun. 2021, 12, 1–7. 

\bibitem{tas18}
F. Tacchino, A. Auffeves, M. F. Santos, and D. Gerace, Steady State Entanglement beyond Thermal Limits. Phys. Rev. Lett. 120.063604 (2018).

\bibitem{msm21}
Gonzalo Manzano, Diego Subero, Olivier Maillet, Rosario Fazio, Jukka P. Pekola, and Édgar Roldán, Thermodynamics of Gambling Demons. Phys. Rev. Lett.126.080603 (2021).

\bibitem{sd11}
Scully MO, Chapin KR, Dorfman KE, Kim MB, Svidzinsky A (2011) Quantum heat engine power can be increased by noise-induced coherence. Proc Natl. Acad Sci USA 108:15097–15100.

\bibitem{gg18}
Arnab Ghosh, David Gelbwaser-Klimovsky, Wolfgang Niedenzu, Alexander I. Lvovsky, Igor Mazets, Marlan O. Scully, and Gershon Kurizki, Two-level masers as heat-to-work converters. Proc. Natl. Acad. Sci. U.S.A. October 2, 2018 115 (40) 9941-9944.

\bibitem{cyc21}
Francesco Caravelli, Bin Yan, Luis Pedro García-Pintos, Alioscia Hamma. (2021) Energy storage and coherence in closed and open quantum batteries. Quantum 5, 505.

\bibitem{mbkb}
Wai-Keong Mok, Kishor Bharti, Leong-Chuan Kwek, Abolfazl Bayat. Optimal probes for global quantum thermometry. Communications Physics volume 4, Article number: 62 (2021).

\bibitem{dv13} 
K. E. Dorfman, D. V. Voronine, S. Mukamel, and M. O. Scully (2013) Photosynthetic reaction center as a quantum heat engine. Proc. Natl. Acad. Sci. U.S.A. 110, 2746 (2013). 

\bibitem{sprl}
M. O. Scully, Quantum Photocell: Using Quantum Coherence to Reduce Radiative Recombination and Increase Efficiency, Phys. Rev. Lett. 104, 207701 (2010).

\bibitem{sz03}
M. O. Scully, M. S. Zubairy, G. S. Agarwal, and H. Walther, Extracting work from a single heat bath via vanishing quantum coherence, Science 299, 862 (2003).

\bibitem{rm12}	
S. Rahav, U. Harbola, and S. Mukamel, Heat fluctuations and coherences in quantum heat engines, Phys. Rev. A 86, 043843 (2012).

\bibitem{lk15}
R. Uzdin, A. Levy, and R. Kosloff, Equivalence of Quantum Heat Machines, and Quantum-Thermodynamic Signatures, Phys. Rev. X 5, 031044 (2015).

\bibitem{osm20}
K. Ono, S.N. Shevchenko, T. Mori, S. Moriyama, and Franco Nori, Analog of a Quantum Heat Engine Using a Single Spin Qubit. Phys. Rev. Lett. 125, 166802 (2020).

\bibitem{qd21} Md Qutubuddin and Konstantin E. Dorfman, Incoherent control of optical signals: Quantum-heat-engine approach, Phys. Rev. Research 3, 023029 – Published 9 April 2021.

\bibitem{sdf13}
Schlawin, F. Dorfman, K. E. Fingerhut, B. P. Mukamel, S. Suppression of population transport and control of exciton distributions by entangled photons. Nat. Commun. 2013, 4, 1-7.

\bibitem{ryangood2020}
Ryan K. BurdickRyan K. Burdick, George C. Schatz, and Theodore Goodson III, Enhancing Entangled Two-Photon Absorption for Picosecond Quantum Spectroscopy. J. Am. Chem. Soc. 2021, 143, 41, 16930–16934.

\bibitem{smuk20road}
Shaul Mukamel, $et~al$, Roadmap on quantum light spectroscopy.  J. Phys. B: At. Mol. Opt. Phys. 53 072002 (2020).


\bibitem{zhpag2021}
Zhedong Zhang, Tao Peng, Xiaoyu Nie, Girish S. Agarwal, and Marlan O. Scully, Entangled Photons Enabled Time- and Frequency-Resolved Coherent Raman Spectroscopy in Condensed Phase Molecules. arXiv:2106.10988v2.

\bibitem{dsm16}
Dorfman, K. E. Schlawin, F. Mukamel, S. Nonlinear optical signals and spectroscopy with quantum light. Rev. Mod. Phys. 2016, 88, 045008.

\bibitem{oleg2020}
Oleg Varnavski and Theodore Goodson III, Two-Photon Fluorescence Microscopy at Extremely Low Excitation Intensity: The Power of Quantum Correlations. J. Am. Chem. Soc. 2020, 142, 12966-12975.

\bibitem{rahmu2010}
S. Rahav and S. Mukamel, Chapter 6 - Ultrafast Nonlinear Optical Signals Viewed from the Molecule's Perspective: Kramers-Heisenberg Transition-Amplitudes versus Susceptibilities, Adv. At. Mol. Opt. Phys. 59, 223 - 263 (2010).

\bibitem{sdm18}
Frank Schlawin, Konstantin E. Dorfman, and Shaul Mukamel, Entangled Two-Photon Absorption Spectroscopy. Acc. Chem. Res. 2018, 51, 2207-2214.
	
\bibitem{kpra18} 
K. E. Dorfman, D. Xu, and J. Cao, Efficiency at maximum power of a laser quantum heat engine enhanced by noise-induced coherence, Phys. Rev. E 97, 042120 (2018).

\bibitem{mcr9}
Abramavicius D, Palmieri B, Voronine DV, Sanda F, Mukamel S (2009) Coherent multidimensional optical spectroscopy of excitons in molecular aggregates; quasiparticle versus supermolecule perspectives. Chem Rev 109(6):2350–2408.

\bibitem{harris1989}
D. C. Harris and M. D. Bertolucci, Symmetry and Spectroscopy, Introduction to Vibrational and Electronic Spectroscopy (Dover, New York, 1989).

\bibitem{klr14} 
R. Kosloff and A. Levy, Quantum heat engines and refrigerators: Continuous devices, Annual Review of Physical Chemistry 65 (2014), pp. 365–393.

\bibitem{fenfmuk2021}
Feng Chen and Shaul Mukamel, Vibrational Hyper-Raman Molecular Spectroscopy with Entangled
Photons. ACS Photonics 2021 8 (9), 2722-2727.

\bibitem{sac21}
Anatoly Svidzinsky , Girish Agarwal, Anton Classen, Alexei V. Sokolov, Aleksei Zheltikov, M. Suhail Zubairy, and Marlan O. Scully, Enhancing stimulated Raman excitation and two-photon absorption using entangled states of light. Phys. Rev. Research.3.043029.


\end{thebibliography}
\end{document}